\documentclass[11pt]{article} 
\pdfoutput=1
\usepackage{amsmath,amssymb,amsfonts,fullpage,graphicx,nicefrac,slashed,tikz,tikzfeynman}
\usepackage[compress]{cite}
\usepackage[normalem]{ulem} 
\usepackage[normal,font=small,labelfont=bf,labelsep=period]{caption}
\renewcommand{\arraystretch}{1.5}
\renewcommand{\hat}{\widehat}
\newcommand{\order}[1]{\mathcal{O}\left({#1}\right)}
\newcommand{\br}{\mathrm{BR}}

\usepackage{color}
\definecolor{colorLink}{rgb}{0.9,0,0} 
\definecolor{colorCite}{rgb}{0,0.7,0} 
\definecolor{colorURL} {rgb}{0,0,0.8} 
\definecolor{colorCapt}{rgb}{0.5,0,1} 
\definecolor{colorPink}{rgb}{1,0,0.7} 
\usepackage[colorlinks=true,linktocpage=true,linkcolor=colorLink,citecolor=colorCite,urlcolor=colorURL]{hyperref}

\title{\vspace{-2cm}
\begin{flushright}{\small EFI-15-23}\vspace{1cm}\end{flushright}
\noindent{\LARGE \bf Composite spin-1 resonances at the LHC}\\}

\author{Matthew Low$^a$, Andrea Tesi$^b$, Lian-Tao Wang$^a$}
\date{\it \small $^a$Department of Physics, Enrico Fermi Institute, and Kavli Institute for\\
Cosmological Physics, University of Chicago, Chicago, IL 60637\\
\small $^b$Department of Physics, Enrico Fermi Institute, University of Chicago, Chicago, IL 60637}

\begin{document}
\begin{titlepage}
\maketitle
\thispagestyle{empty}

\begin{abstract}
\noindent
In this paper, we discuss the signal of composite spin-1 resonances at the LHC.  Motivated by the possible observation of a diboson resonance in the 8 TeV LHC data, we demonstrate that vector resonances from composite Higgs models are able to describe the data.  We pay particular attention to the role played by fermion partial compositeness, which is a common feature in composite Higgs models.  The parameter space that is both able to account for the diboson excess and passes electroweak precision and flavor tests is explored.  Finally, we make projections for signals of such resonances at the 13 TeV run of the LHC.
\end{abstract}

\vfill
\noindent\line(1,0){188}\\\medskip
\footnotesize{E-mail: \texttt{\href{mailto:mattlow@uchicago.edu}{mattlow@uchicago.edu}, \href{mailtoatesi@uchicago.edu}{atesi@uchicago.edu}, \href{mailto:liantaow@uchicago.edu}{liantaow@uchicago.edu}}}
\end{titlepage}

\tableofcontents

\section{Introduction}\label{sec:intro}

The hunt for new physics will start again with the second run of the LHC.  From the 7 and 8 TeV data collected by the ATLAS and CMS experiments no evident signals of new physics have emerged thus far.  For many models and frameworks, this makes the 13 TeV LHC run the last place to probe the natural region of their parameter spaces.  In the current LHC data, however, there are some anomalies at the $\sim 3\sigma$ level.  While $3\sigma$ observations could turn out to be statistical fluctuations, it can be useful to take some of them seriously and explore their consequences within a given model.  

In this paper we take motivation from the observation of an excess in the ATLAS hadronic diboson search~\cite{Aad:2015owa} (and CMS too~\cite{Khachatryan:2014hpa}).  This search looks for resonances decaying to a pair of boson tagged jets.  ATLAS has searches in the $WZ$, $WW$, and $ZZ$ channels which differ by the jet mass selection applied and observes local excesses of 3.4$\sigma$, 2.6$\sigma$, and 2.9$\sigma$, respectively.

As a guideline, the excess requires the resonance to have a sizable coupling to vector bosons and a sufficiently large production cross-section to be observed, $\simeq 5-10$ fb.  The most natural interpretation of this excess is a spin-1 resonance (or more precisely a multiplet of nearly degenerate spin-1 resonances) of about 2 TeV with a large coupling to vector bosons (to dominate the branching ratio) and a smaller coupling to quarks (for Drell-Yan production).  Crucially, a large coupling to leptons must be avoided as bounds from dilepton searches are quite constraining~\cite{Aad:2014cka,Khachatryan:2014fba}.  Models with these features have been the subject of most of the recent papers on the subject including technicolor models~\cite{Fukano:2015hga}, effective models with spin-1 resonances \cite{Franzosi:2015zra}, left-right symmetric models~\cite{Dobrescu:2015qna,Gao:2015irw,Brehmer:2015cia}, (composite) SU(2) triplet models~\cite{Pappadopulo:2014qza, Aguilar-Saavedra:2015rna,Thamm:2015csa,Sanz:2015zha,Bian:2015ota}, and non-custodial models~\cite{Carmona:2015xaa}.  See~\cite{Hisano:2015gna,Cheung:2015nha,Xue:2015wha,Chao:2015eea,Omura:2015nwa,Chen:2015xql,Chiang:2015lqa,Cacciapaglia:2015nga,Alves:2015mua,Cao:2015lia,Abe:2015jra,Anchordoqui:2015uea,Englert:2015oga} for other models.

\paragraph{}
In this work we are interested in interpreting the excess in the context of composite Higgs models with partially composite vectors and fermions (see~\cite{Contino:2010rs,Panico:2015jxa} for reviews).  This framework consists of two sectors.  One sector, called the ``composite sector,'' contains the Higgs multiplet as the Goldstone bosons of a symmetry that is spontaneously broken at a scale $f$.  The Higgs interacts with resonances in the strong sector with a large coupling $g_\rho$.  The second sector, called the ``elementary sector,'' contains the standard model gauge and fermion fields.\footnote{Note that the actual standard model fields are linear combinations of the elementary and composite fields.}

The two sectors communicate via a linear mixing between fields of the same spin.  This mixing breaks the global symmetries and generates a potential for the Goldstone Higgs that triggers electroweak symmetry breaking (EWSB) at a scale $v<f$.  The mixing angles between the composite and elementary fields ({\it i.e.} the degree of compositeness) are proportional to the ratios of standard model couplings $g_{\rm SM}$ (where $g_{\rm SM}$ is representative of gauge couplings or Yukawa couplings)  and the composite sector ones $g_\rho$ as required to reproduce the standard model.  Therefore, in this picture the resonances of the strong sector, in particular, the spin-1 fields, couple to standard model fermions and transversely polarized gauge bosons with a coupling that is naturally suppressed by $\sim g_{\rm SM}^2 / g_\rho$.  Given that the Higgs is part of the strong sector, however, the resonances couple strongly with the longitudinally polarized components of the $W$ and the $Z$.

These facts make composite Higgs models a particularly attractive framework in which to study diboson production.  In these models, the expected mass of the lightest resonance is $m_\rho \sim g_\rho f$.  The scale $f$ is bounded from below around 600 GeV from Higgs measurements~\cite{ATLAS:2015bea,Khachatryan:2014jba} which means for a 2 TeV resonance, $g_\rho$ is bounded from above at roughly $g_\rho \lesssim 4-5$.  Being forced to use a moderate value of $g_\rho$ means that $g_{\rm SM}^2 / g_\rho$ is not overly suppressed and can produce vector resonances with the appropriate rate.

It should be emphasized that in this class of models, the standard model fermions are linearly coupled to composite fermions (which in turn are tightly coupled to the spin-1 resonances)~\cite{Kaplan:1991dc}.  This introduces, via mixing, an additional sizable contribution to the coupling of the vector resonances to standard model quarks (see also~\cite{Bian:2015ota} in relation to the diboson excess).  One challenge in this framework of partially composite fermions is that light colored top partners, the fermions which mix  with the standard model top quark, are typically expected to be close to the TeV scale due to Higgs mass considerations \cite{Contino:2006qr}. When kinematically open, decays of vectors into fermion partners are typically dominant~\cite{Barducci:2012kk,Bellazzini:2012tv,Vignaroli:2014bpa,Greco:2014aza}.\footnote{We work in four dimensions, but studies in five dimensions are equivalent, {\it e.g.} see~\cite{Agashe:2007ki,Agashe:2008jb}.}  In order to suppress this decay, one has to assume a large mass scale for the composite fermions, or, in other words, a small elementary-composite mixing for the top.  Interestingly, this can be naturally achieved in composite twin Higgs models~\cite{Geller:2014kta, Barbieri:2015lqa,Low:2015nqa}.  Given that spin-1 resonances of $2-3$ TeV are expected, in particular, in the scenario of~\cite{Low:2015nqa}, this makes the diboson signal even more prominent in the composite twin Higgs models and substantially differentiates the phenomenology of spin-1 resonances in standard composite Higgs and composite twin Higgs.  The absence of light top-like composite fermions makes spin-1 resonances likely the first signal of composite twin Higgs models.

\paragraph{}
In this paper, we give a detailed discussion on the LHC signals of composite vector resonances.  We discuss production rates and decay channels with an emphasis on the diboson channel.  We also take into account all constraints including precision measurements from LEP, flavor constraints, and direct searches at the LHC.  Particular attention is paid to the ramifications of a vector resonance around 2 TeV.

The rest of the paper is organized as follows. In Sec.~\ref{sec:model} we recall the basic framework with an emphasis on the couplings of the composite vectors to standard model fields and  we introduce the various scenarios we explore.  In Sec.~\ref{sec:lagrangian} we present a simplified Lagrangian and we compute the relevant branching ratios.  Section~\ref{sec:8tev} is devoted to the analysis of the diboson data and the predictions of several benchmarks.  In the same section we point out the relevant constraints on the picture that come from other measurements.  In Sec.~\ref{sec:13tev} we show the rates for the 2 TeV signal at 13 TeV.  Finally, we conclude in Sec.~\ref{sec:conclusions}.  Technical details on the model used are presented in App.~\ref{app:two-site}.

\section{The basic framework}\label{sec:model}

In this section we introduce the various aspects that compromise a composite Higgs model.

\subsection{Composite Higgs overview}

The Higgs is a pseudo Nambu Goldstone boson (pNGB) from the breaking of a global symmetry at a scale $f$, where $f>v$.  The minimal model is SO(5)/SO(4)~\cite{Agashe:2004rs,Contino:2006qr}.  The simplest realizations can be described in four dimensions as a two or three-site model~\cite{Panico:2011pw,DeCurtis:2011yx}, while a more general effective description can be parametrized by the CCWZ formalism~\cite{Coleman:1969sm,Callan:1969sn} (for examples, see~\cite{Contino:2011np,DeSimone:2012fs}).

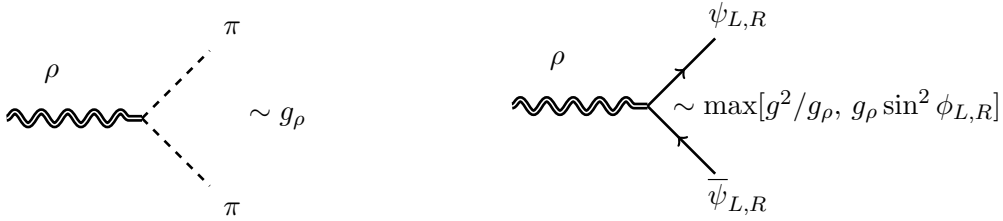
\begin{figure}[h]
\begin{center}
  \begin{tikzpicture}[line width=1pt, scale=.6]
  \draw[vector, double] (-3,0) --(0,0);
  \draw[dashed, line width=1pt] (0,0) --(1.5,1.5);
  \draw[dashed, line width=1pt] (0,0) --(1.5,-1.5);
  \node at (-2,1) {$\rho$};
  \node at (2,2)  {$\pi$};
  \node at (2,-2) {$\pi$};
  \node at (3,0)  {$\sim g_\rho$};
  \end{tikzpicture}
  \hspace{6em}
  \begin{tikzpicture}[line width=1pt, scale=.6]
  \draw[vector, double] (-3,0) --(0,0);
  \draw[fermion, line width=1pt] (0,0) --(1.5,1.5);
  \draw[fermion, line width=1pt] (1.5,-1.5) --(0,0);
  \node at (-2,1) {$\rho$};
  \node at (2,2)    {$\psi_{L,R}$};
  \node at (2,-2)   {$\overline{\psi}_{L,R}$};
  \node at (4.2,0)  {$\sim \mathrm{max}[g^2/g_\rho,\, g_\rho \sin^2\phi_{L,R}]$};
  \end{tikzpicture}
  \caption{Couplings between composite vectors $\rho$ and the longitudinal components of standard model vectors (left) and standard model fermions (right).}
  \label{fig:rhoToVectors}
\end{center}
\end{figure}

The low energy resonances are vectors broadly characterized by a mass $m_\rho\sim g_\rho f$ \cite{Giudice:2007fh}, where $g_\rho$ characterizes the strength of the interactions between particles in the composite sector such as the longitudinal modes $W_L^{\pm}$, $Z_L$, and the Higgs (as shown in Fig.~\ref{fig:rhoToVectors}) and is considered large as it comes from a strongly interacting sector.  All the other resonances  are expected at a scale $m_* = g_* f$ with $g_\rho < g_* \lesssim 4\pi$.  

For simplicity, we consider a single multiplet of vector resonances to be lower than the rest of the compositeness scale as this is the best description of the diboson excess (and actually closely corresponds to the $\rho$ in QCD).\footnote{The unitarization implications of such a resonance have been explored, for example, in~\cite{Falkowski:2011ua,Contino:2011np}.} With the symmetry breaking pattern SO(5)/SO(4), the lowest lying vector modes are in the adjoint of SO(4) which is the \textbf{6}.  In order to reproduce the fermion Yukawa couplings, an additional U(1)$_X$ is required such that the unbroken global symmetry is SO(4) $\times$ U(1)$_X$ = SU(2)$_L$ $\times$ SU(2)$_R$ $\times$ U(1)$_X$, however, the \textbf{6} multiplet is not charged under U(1)$_X$.  The standard model then gauges SU(2)$_L$ and the combination of $T_R^3 + X$ as hypercharge.  The vectors then decompose into SU(2)$_L$ $\times$ U(1)$_Y$ multiplets as
\begin{equation}
  \mathbf{6} \to \mathbf{3}_0 + \mathbf{1}_{0} + \mathbf{1}_\pm,
\end{equation}
where the subscript indicates the hypercharge.  The SU(2)$_L$ triplet $\mathbf{3}_0$ corresponds to both neutral and charged states which we label $\rho^0$ and $\rho^\pm$, respectively (see~\cite{Pappadopulo:2014qza} for a study of the phenomenology).  The masses of the vectors are degenerate and are only split by hypercharge effects, after electroweak symmetry breaking, on the order of $(g'/g_\rho)^2$.

The other vectors $\mathbf{1}_0$ and $\mathbf{1}_\pm$, neutral under SU(2)$_L$, are also present and are approximately mass degenerate with the triplet when SO(4) is unbroken in the strong sector.  We label these states as $\rho_B^0$ and $\rho_C^\pm$, respectively.  In simplified discussions they are often omitted because their interactions with the standard model are subleading in $g'/g$.  We include them in our discussion for completeness.

Unlike the interactions between the composite vectors and standard model (longitudinal) vectors, the interactions between composite vectors and standard model fermions do not originate purely in the strong sector and proceed through the mixing between the composite and elementary states, like vector meson dominance in QCD.  The vector mixing is of order $g/g_\rho$ which induces a coupling of $g^2/g_\rho$ between the composite vectors and standard model fermions (as shown in Fig.~\ref{fig:rhoToVectors}).

In the standard picture of compositeness this is not the only contribution to the composite vector coupling with standard model fermions; there is also a contribution from the partial compositeness of the standard model quarks.  This is a mechanism to give mass to chiral fermions and is a linear mixing between standard model elementary quarks and the composite vector-like fermions.

For the sake of our discussion, it is only important to notice that the chiral standard model fermions are given by a linear combination of an elementary and composite state with mixing given by the angle $\sin\phi_{L,R}^f$ where the species label $f$ allows for each fermion to have a different degree of compositeness (we will frequently use the shorthand $s_{L,f} = \sin\phi_L^f$ and $c_{L,f} = \cos\phi_L^f$).  The mixings are then constrained to reproduce the correct Yukawa couplings
\begin{equation} \label{eq:yukawa}
  y_f = \frac{m_\Psi}{f} \sin\phi_L^f \sin\phi_R^f .
\end{equation}
Above, $m_\Psi$ is the characteristic mass of the composite fermions.  Thus, the coupling between standard model fermions and vector resonances receives contributions both from vector mixing and from fermion mixing, with the fermion mixing contribution proportional to $g_\rho \sin^2\phi_{L,R}$, depending on the chirality of the current.  

These are summarized pictorially in Fig.~\ref{fig:rhoToVectors}.

\subsection{Summary of the interactions and benchmark models}\label{sec:benchmarks}

The qualitative description above is summarized quantitatively in Table~\ref{tab:couplings}.  The table shows that vector resonances couple to standard model vectors with a strength $g_\rho$.  The couplings to standard model fermions, however, are more complicated.  Starting with the $\rho^{0,\pm}$ we see that it only couples to the left handed currents.  The interactions with quarks have a term $g^2/g_\rho$ from vector mixing and $g_\rho s_{L,q}^2$ from fermion mixing.

In the couplings of vector resonances to the left handed currents we include an extra parameter $a_L$, which is common to different generations.  In the concrete two-site model (reviewed both in App.~\ref{app:two-site} and more fully in the appendix of~\cite{Panico:2011pw}), $a_L = 1$. However, more generally,   it is a free parameter in the CCWZ parametrization (for example, see~\cite{Greco:2014aza}).  By default we will discuss the two-site case when $a_L = 1$, but in certain cases we will present results for the flipped sign $a_L = -1$ case.  As can be seen from Table~\ref{tab:couplings}, choosing $a_L = -1$ can avoid cancellations in the resonance coupling to quarks which would otherwise lead to very small rates.

\begin{table}[h!]
\begin{center}
\begin{tabular}{ c || c | c c c | c c }
  & $VV$, $Vh$
  & $\bar{q}_L \gamma^\mu q_L$         & $\bar{u}_R \gamma^\mu u_R$        & $\bar{d}_R \gamma^\mu d_R$ 
  & $\bar{\ell}_L \gamma^\mu \ell_L$   & $\bar{e}_R \gamma^\mu e_R$ \\ \hline \hline
  $\rho^{0,\pm}$ & $g_\rho$ 
                 & $\displaystyle-\frac{g^2}{g_\rho}(1 - a_L \frac{g^2_\rho}{g^2} s_{L,q}^2)\tau^a$ 
                 & -- & -- & $-\displaystyle\frac{g^2}{g_\rho}\tau^a$ & -- \\
  $\rho_B^0$   & $g_\rho$ 
               & $-\displaystyle\frac{1}{6}\frac{g'^2}{g_\rho}(1 + 3 a_L \frac{g_\rho^2}{g'^2} s_{L,q}^2)$ 
               & $-\displaystyle\frac{2}{3}\frac{g'^2}{g_\rho}$
               & $\displaystyle\frac{1}{3}\frac{g'^2}{g_\rho} $ & $\displaystyle\frac{1}{2}\frac{g'^2}{g_\rho}$
               & $\displaystyle\frac{g'^2}{g_\rho}$ \\
 $\rho_C^\pm$ & $g_\rho$ & -- & -- & -- & -- & --
 \end{tabular}
  \caption{Summary of SU(2)$_L \times$ U(1)$_Y$ invariant couplings between vector resonances and standard model fermions $q_L$, $u_R$, and $d_R$, massive gauge bosons $V$, and Higgs boson $h$  at leading order in $g/g_\rho$ (and $g'/g_\rho$).}
  \label{tab:couplings}
\end{center}
\end{table}

Regarding the expected size of the mixing angles, only the top must have a sizable degree of compositeness.  The reason is that in standard composite Higgs, light top partners are required ({\it i.e.} $m_\Psi \simeq f$) in order to achieve the observed Higgs mass which leads to $\sin\phi^t_{L,R} \sim 1$ according to Eq.~\eqref{eq:yukawa}.  While it is conceivable that the top mixings can be made smaller at the price of tuning, it is interesting to note that at least the top left mixing can be naturally small in the composite twin Higgs scenario.  The other quarks usually have small mixing angles. In this paper, we explore several limits, paying attention to possible precision constraints.  We omit any lepton mixing in the table as we treat them as elementary.

It is interesting to note that  the mixing of the left handed fermions  has a much larger effect than that of the right handed fermions because the vector phenomenology is primarily determined by the $\rho^{0,\pm}$.  The right handed fermions only couple with the SU(2)$_L$ singlet $\rho_B^0$.  Moreover, given the composite scenario under consideration, there is no dependence on the right handed mixing angles (see App.~\ref{app:two-site}).  This observation allows us to straightforwardly present the impact of fermion partial compositeness as a function of $s_L$.

To keep the discussion simple and to best fit the current experimental data, we do not consider the possibility of the vector resonances decaying to fermions in the composite sector.  While such decays are generic~\cite{Barducci:2012kk,Bellazzini:2012tv,Vignaroli:2014bpa,Greco:2014aza}, especially when $m_\Psi \sim f$, the phenomenological consequences are beyond the scope of this work.  An example of a concrete natural model where such decays are not expected is the twin composite Higgs model where the Higgs mass can be correctly achieved without light top partners.

As a final remark on the table, we only list interactions that are SU(2)$_L$ $\times$ U(1)$_Y$ invariant, \textit{i.e} ignoring the effects of electroweak symmetry breaking.  EWSB will generate additional interactions, even where there are blanks in the table, but such interactions are expected to be small corrections compared to the couplings and masses of the composite sector.  In particular, we find the corrections to be negligible for the interactions between composite vectors and standard model quarks.

\paragraph{}
The couplings in Table~\ref{tab:couplings} are rather simple, but still they provide a rich  spectrum of phenomenological possibilities. In order to simplify the discussion we will focus in the following on the scenarios:
\begin{itemize}

  \item[1.] {\bf Elementary fermions}: Fermions are taken to be elementary and only couple to vector resonances from vector mixing.  In our notation, this amounts to setting all of the fermion mixing angles to zero, $s_{L,f} = s_{R,f} = 0$.  In this case the only free parameters are $g_\rho$ and $m_\rho$. Despite the fact that this scenario corresponds to the limit of massless quarks (see Eq.~\eqref{eq:yukawa}), we still consider it as a possible benchmark (along the lines of \cite{Pappadopulo:2014qza,Thamm:2015csa}) since it allows for a simple discussion of the relevant constraints.
  
  \item[2.] {\bf Composite top}: In this case we consider the $(t_L,b_L)$ doublet to have a sizable degree of compositeness (as well as the $t_R$).  Here the relevant parameters are $g_\rho$, $m_\rho$, and $s_{L,t}$.  Given that in this paper we do not impose the constraint of the Higgs mass in the parameter space, we consider $s_{L,t}$ over its full range, despite the fact that in concrete realizations extreme values such as very close to zero or one are unlikely.  Moreover, $s_{L,t} \simeq 1$ can be compatible with data only if there is a flavor symmetry at work.  As a benchmark, we always assume a U(2)$^2$ flavor symmetry in the left handed mixings~\cite{Barbieri:2012tu,Matsedonskyi:2014iha}.

  \item[3.] {\bf Composite quarks}: In this case we allow for the lighter standard model quarks ($u$, $d$, $s$, and $c$) to be partially composite.  We choose two benchmark values for the left compositeness of the top of
\begin{subequations}\label{eq:benchmarks-compQuark}
\begin{align}
  \sin\phi_L^t &= 0.4,  &\text{standard composite Higgs} \\
  \sin\phi_L^t &= 0.1,  &\text{composite twin Higgs}
\end{align}
\end{subequations}
where the composite twin Higgs scenario has a smaller $\sin\phi_L^t$ because the twin mechanism naturally allows for $m_\Psi \gg f$.   The parameters are $g_\rho$, $m_\rho$, and $s_{L,q}$, where $s_{L,q}$ is the left compositeness of the lighter quarks and is taken to be the same for the first two generations ({\it i.e.} we have in mind an underlying U(2)$^2$ flavor symmetry).

\end{itemize}

In addition to these classes, we also look at two possibilities of the relation between $m_\rho$ and $g_\rho f$.  We define the parameter $c_H$ as
\begin{equation}\label{eq:ch-definition}
  c_H = \frac{m_\rho^2}{g_\rho^2 f^2}.
\end{equation}
In the two-site model we have that $c_H = 1/2$, but we also consider the case when $c_H = 1$.

\section{Effective description}\label{sec:lagrangian}

In this section we review the interactions between the composite vector and both standard model fermions and standard model vectors.  The full Lagrangian is shown in App.~\ref{app:two-site}.  Additionally,  while the discussion in this section takes place in the electroweak symmetric limit for simplicity, all numerical results presented in this work use the appropriate equations after electroweak symmetry breaking.

\subsection{Low energy interactions}

The form of the interaction between the $\rho$ triplet and fermions is
\begin{equation}
\mathcal{L} \supset
 -\left(\frac{g^2}{g_\rho} - a_L g_\rho s_{L,f}^2\right) \rho_\mu^a J^{\mu a},
\end{equation}
where the left compositeness $s_{L,f}$ is different for each type of fermion $f$.  We assume elementary leptons and a U(2) flavor symmetry which means we have only two parameters $s_{L,t}$ controlling the third generation left compositeness and $s_{L,q}$ which controls the lighter quarks.  For leptons $s_L = 0$.  Note that the current only includes left handed fermions
\begin{equation}
 J^{\mu a} = \sum_f \bar{f}_L \gamma^\mu \tau^a f_L .
\end{equation}
Standard model fermions do not couple to $\rho^\pm_C$, while both the left and right currents couple to the $\rho_B$ with couplings that can be read from Table~\ref{tab:couplings}.

The interaction between the $\rho$ and standard model vectors comes from mixing due to electroweak symmetry breaking.  It is simpler to see, however, through the interaction between the $\rho$ and the Goldstone modes from the operator
\begin{equation}
\mathcal{L} \supset i g_\rho c_H \rho_\mu^a (H^\dagger \tau^a D^\mu H - (D^\mu H)^\dagger \tau^a H).
\end{equation}
The Higgs doublet $H$ contains the physical Higgs $h$, but also the Goldstone modes of the $W^\pm$ and $Z$, $\pi^\pm$ and $\pi^0$, respectively.
\begin{equation}
H = \left( \pi^+ , \; \frac{1}{\sqrt{2}}(v + h + i \pi^0) \right).
\end{equation}
As shown in Table~\ref{tab:couplings} the strength of the coupling is $g_\rho$.

\subsection{Production rates and branching ratios}

There are two production mechanisms for the vector resonances: Drell-Yan and vector boson fusion. Vector boson fusion is very subdominant even for large $g_\rho$.  We neglect the contribution of vector boson fusion to the total rate (though it could offer a useful cross-check, albeit needing a quite large integrated luminosity).

The production cross-section depends on the $\rho$ coupling to light quarks and goes like
\begin{equation}
  \sigma(\rho) \sim \left(\frac{g^2}{g_\rho} - a_L g_\rho s_{L,q}^2\right)^2.
\end{equation}
From this it is straightforward to see that when $s_{L,q} = 0$, increasing $g_\rho$ will decrease the overall rate.  For sufficiently large quark compositeness the behavior reverses and starts to grow with $g_\rho$.

The branching ratios can be approximated, in the $m_\rho \gg m_t$ limit, as
\begin{subequations}\label{eq:brs}
\begin{align}
  \br(\rho^\pm \to W^\pm Z/h)
                &= \frac{1}{2}\frac{c_H^2}{c_H^2 + 6 a_L^2 s_{L,t}^4 + 12 a_L^2 s_{L,q}^4 - a_L (12 s_{L,t}^2 + 24 s_{L,q}^2)(\nicefrac{g}{g_\rho})^2 + 24 (\nicefrac{g}{g_\rho})^4 }, \\
  \br(\rho^\pm \to \ell^\pm \nu)
                &= \frac{4 (\nicefrac{g}{g_\rho})^4}{c_H^2 + 6 a_L^2 s_{L,t}^4 + 12 a_L^2 s_{L,q}^4 - a_L (12 s_{L,t}^2 + 24 s_{L,q}^2)(\nicefrac{g}{g_\rho})^2 + 24 (\nicefrac{g}{g_\rho})^4 }, \\
  \br(\rho^\pm \to t\bar{b})
                &= \frac{6 a_L^2 s_{L,t}^4 - 12 a_L s_{L,t}^2 (\nicefrac{g}{g_\rho})^2 + 6 (\nicefrac{g}{g_\rho})^4}{c_H^2 + 6 a_L^2 s_{L,t}^4 + 12 a_L^2 s_{L,q}^4 - a_L (12 s_{L,t}^2 + 24 s_{L,q}^2)(\nicefrac{g}{g_\rho})^2 + 24 (\nicefrac{g}{g_\rho})^4 }, \\
  \br(\rho^\pm \to jj)
                &= \frac{12 a_L^2 s_{L,q}^4 - 24 a_L s_{L,q}^2 (\nicefrac{g}{g_\rho})^2 + 12 (\nicefrac{g}{g_\rho})^4}{c_H^2 + 6 a_L^2 s_{L,t}^4 + 12 a_L^2 s_{L,q}^4 - a_L (12 s_{L,t}^2 + 24 s_{L,q}^2)(\nicefrac{g}{g_\rho})^2 + 24 (\nicefrac{g}{g_\rho})^4 },
\end{align}
\end{subequations}
where $\ell=(e,\mu)$ and $j=(u,d,c,s)$.  The branching ratios of $\rho^\pm$ and $\rho^0$ are simply related by
\begin{subequations}\label{eq:br-corr}
\begin{align}
  \br(\rho^\pm \to W^\pm Z) &= \br(\rho^0 \to W^+ W^-) = \br(\rho^\pm \to W^\pm h) = \br(\rho^0 \to Z h), \\
  \br(\rho^\pm \to \ell^\pm \nu) &= 2\br(\rho^0 \to \ell^+ \ell^-) , \\
  \br(\rho^\pm \to t\bar{b}) &= 2\br(\rho^0 \to t\bar{t})= 2 \br(\rho^0 \to b\bar{b}) , \\
  \br(\rho^\pm \to jj) &= \br(\rho^0 \to jj) .
\end{align}
\end{subequations}
These formulae are useful to understand the dominance of some decay channels over others.  For example, for elementary fermions we see that for $g_\rho \gtrsim g$ the diboson branching ratio becomes $\br(\rho \to VV) \simeq 0.5$ independent of $c_H$.  The fermion branching ratios fall quickly as $\br(\rho \to f\bar{f}) \sim g^4 / g_\rho^4$.

The branching ratios of $\rho_B$ will be similar to the ones of $\rho^0$. The main difference is in the decays to quarks which depend on the hypercharge and include both chiralities.

It is useful to compare ratios of branching ratios to gain intuition as to which other direct searches can be sensitive to these scenarios.
%
\begin{subequations}\label{eq:br-ratios}
\begin{align}
\frac{\br(\rho^0 \to \ell^+ \ell^-)}{\br(\rho^0 \to W^+ W^-)} &= \frac{4}{c_H^2} \frac{g^4}{g_\rho^4}, \\
\frac{\br(\rho^0 \to t \bar{t} )}{\br(\rho^0 \to W^+ W^-)}    &= a_L^2 \frac{6 s_{L,t}^4}{c_H^2} - a_L \frac{12 s_{L,t}^2}{c_H^2} \frac{g^2}{g_\rho^2}+ \frac{6}{c_H^2} \frac{g^4}{g_\rho^4}, \\
\frac{\br(\rho^0 \to j j)}{\br(\rho^0 \to W^+ W^-)}           &= a_L^2 \frac{24 s_{L,q}^4}{c_H^2} - a_L \frac{48 s_{L,q}^2}{c_H^2} \frac{g^2}{g_\rho^2}+ \frac{24}{c_H^2} \frac{g^4}{g_\rho^4}.
\end{align}
\end{subequations}
For $g_\rho \simeq g$ one can see that dilepton and single lepton with missing energy searches can be constraining.  We also see that for large enough $s_{L,t}$ constraints from $t\bar{t}$ and $t\bar{b}$ searches are relevant.  While this suggests that dijet constraints can be relevant for larger $s_{L,q}$,  we will see that precision electroweak constraints are much stronger than dijet searches.

\begin{figure}
\begin{center}
\includegraphics[width=0.45\textwidth]{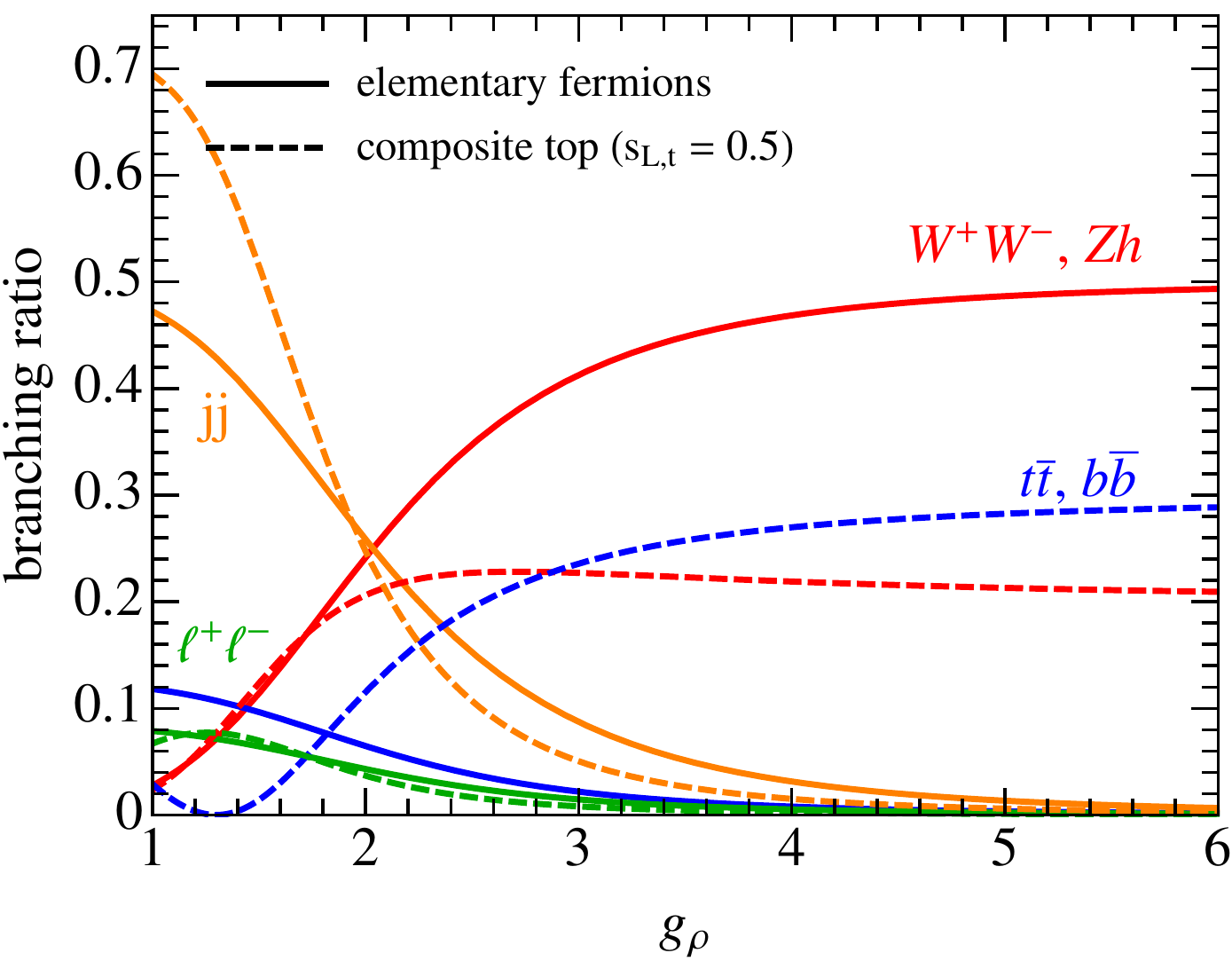} \quad\quad
\includegraphics[width=0.45\textwidth]{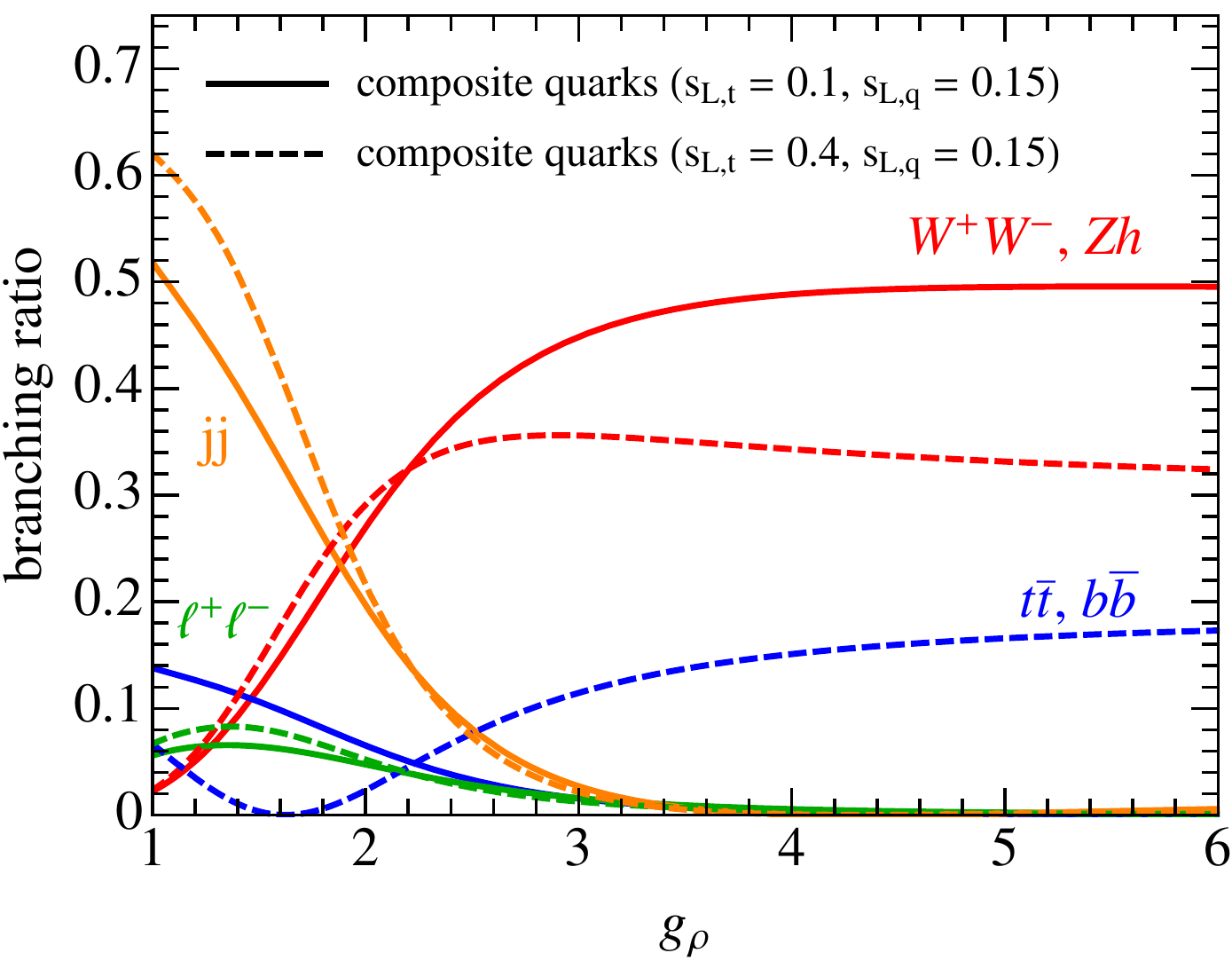}
\caption{Branching ratios for the $\rho^0$. These are plotted for $m_\rho =$ 2 TeV, but the variation with mass is negligible.  The branching ratios for $\rho^\pm$ are correlated to those of $\rho^0$ as shown in Eq.~\eqref{eq:br-corr}. In particular, $\br(W^\pm Z) = \br(W^+ W^-)=\br(Z h)=\br(W^+ h)$,  $\br(t\bar{b}) = 2 \br(t\bar{t}) = 2\br(b\bar{b})$, and $\br(\ell^\pm \nu) = 2 \br(\ell^+ \ell^-)$.  On the left we show elementary fermions (solid) and a composite top with $s_{L,t} = 0.5$ (dashed). On the right we show $s_{L,q}=0.15$ with $s_{L,t}=0.4$ (solid) and $s_{L,q}=0.15$ with $s_{L,t}=0.1$ (dashed) corresponding to the standard composite Higgs and composite twin Higgs benchmarks, respectively.}
\label{fig:branchingRatios}
\end{center}
\end{figure}

Figure~\ref{fig:branchingRatios} contrasts the branching ratios for the different scenarios we have outlined in Sec.~\ref{sec:model} to provide some intuition into the results in Sec.~\ref{sec:8tev}.

\section{Possible signals in 8 TeV LHC data}\label{sec:8tev}

Recently, ATLAS has reported an excess of 3.4$\sigma$ in the $WZ$ channel of a boson tagged dijet search~\cite{Aad:2015owa}.  The related channels $WW$ and $ZZ$, differing by the jet mass selection, accordingly found excesses of 2.6$\sigma$ and 2.9$\sigma$, respectively.  Due to the difficulty of distinguishing hadronically decaying $W$'s and $Z$'s these channels are correlated and the 2 TeV resonance could conceivably be neutral or charged or a multiplet of states as we consider.

To crudely estimate the total rate required to reproduce the excess, we compare the number of simulated $W'$ signal events appearing in the dijet mass distribution of~\cite{Aad:2015owa} with the $W'$ cross-section, and then compare this to the number of excess events in the dijet mass distribution.  We find in the $WZ$ channel this corresponds to a signal of $\sim$ 10 fb.  Of course, it is certainly possible that this is an upward fluctuation.  Therefore, a cross-section of
\begin{equation}
\sigma \times \br(VV) \simeq 5 - 10~\mathrm{fb},
\end{equation}
would account for the excess.

The estimate above applies to a single (charged) spin-1 vector. However, in the case that the resonance has a neutral state degenerate in mass (such as the case  in the present scenario), given the poor efficiency for differentiating the $W$ and $Z$ fatjets, the neutral state which decays to $WW$ can also contribute to the signal events (notice also that the hadronic branching fractions of $W$ and $Z$ are very similar).  Indeed, given that $7-8$ events (with 20 fb$^{-1}$) are observed in the 3 bins closest to 2 TeV~\cite{Aad:2015owa}, with an acceptance $\mathcal{A}$ of roughly 0.2~\cite{Aad:2015owa} and estimated efficiencies $\epsilon_{WZ} \sim 50\%$ and $\epsilon_{WW} \sim 50\%$ for $WZ$ and $WW$ reconstruction, respectively, we find
\begin{equation}
(\mathcal{A} \times \epsilon_{WZ}) \sigma(\rho^\pm) \times \br(WZ\to\mathrm{hadrons})
+ (\mathcal{A} \times \epsilon_{WW}) \sigma(\rho^0) \times \br(WW\to\mathrm{hadrons}) 
= \frac{7-8~\mathrm{events}}{20~\mathrm{fb}^{-1}}
\end{equation}
which gives again an estimate of $\simeq 5-10$ fb for the total cross-section (see~\cite{Allanach:2015hba} for more careful estimates of the efficiencies). 

Both ATLAS and CMS have performed other searches for diboson resonances in the semileptonic~\cite{Khachatryan:2014gha,Aad:2014xka,Aad:2015ufa} and fully leptonic channels~\cite{Aad:2014pha,Khachatryan:2014xja}.  Due to a smaller branching ratio of $W$'s and $Z$'s to leptons, these searches are not quite as strong and yield bounds on $W'$ bosons from extended gauge models of $m_{W'} \leq 1.5$ TeV.  Other searches like those for $WH$ or $ZH$ resonances can also be relevant given that the branching ratio for these decays is $\simeq 50\%$.  Searches by CMS include the final states $Z(J)H(\tau^+\tau^-)$~\cite{Khachatryan:2015ywa}, $V(jj)H(\text{jets})$~\cite{Khachatryan:2015bma}, and $W(\ell^\pm\nu)H(b\bar{b})$~\cite{CMS:2015gla}, while ATLAS has a search for $Z(\ell^+\ell^-\text{ or }\nu\bar{\nu})H(b\bar{b})$ and $W(\ell^\pm\nu)H(b\bar{b})$~\cite{Aad:2015yza}.  Only the $V(jj)H(\text{jets})$ search has relevant sensitivity and comes just short of excluding the elementary fermions case.  We do not include these limits in our results, but acknowledge that in cases where fermion compositeness increases the diboson rate, there is tension at 1$\sigma$ with the $VH$ searches.

\paragraph{}
In the following where we compute cross-sections, we use Madgraph 5 v.2.3.0~\cite{Alwall:2014hca}.  From Madgraph we find a $k$ factor of $k=1.4$ both at 8 TeV and 13 TeV which we apply to all cross-sections.

\subsection{Model interpretations}

In this section, we systematically walk through each scenario of fermion compositeness and discuss which cases produce appropriately large diboson rates for a 2 TeV resonance while simultaneously satisfying both direct searches and indirect tests.  First, the results will be shown, then in Sec.~\ref{sec:8tevdirect} the direct constraints will be discussed and in Sec.~\ref{sec:8tevindirect} the indirect constraints will be discussed.

\subsubsection*{Elementary fermions} 

\begin{figure}
\begin{center}
\includegraphics[width=0.5\textwidth]{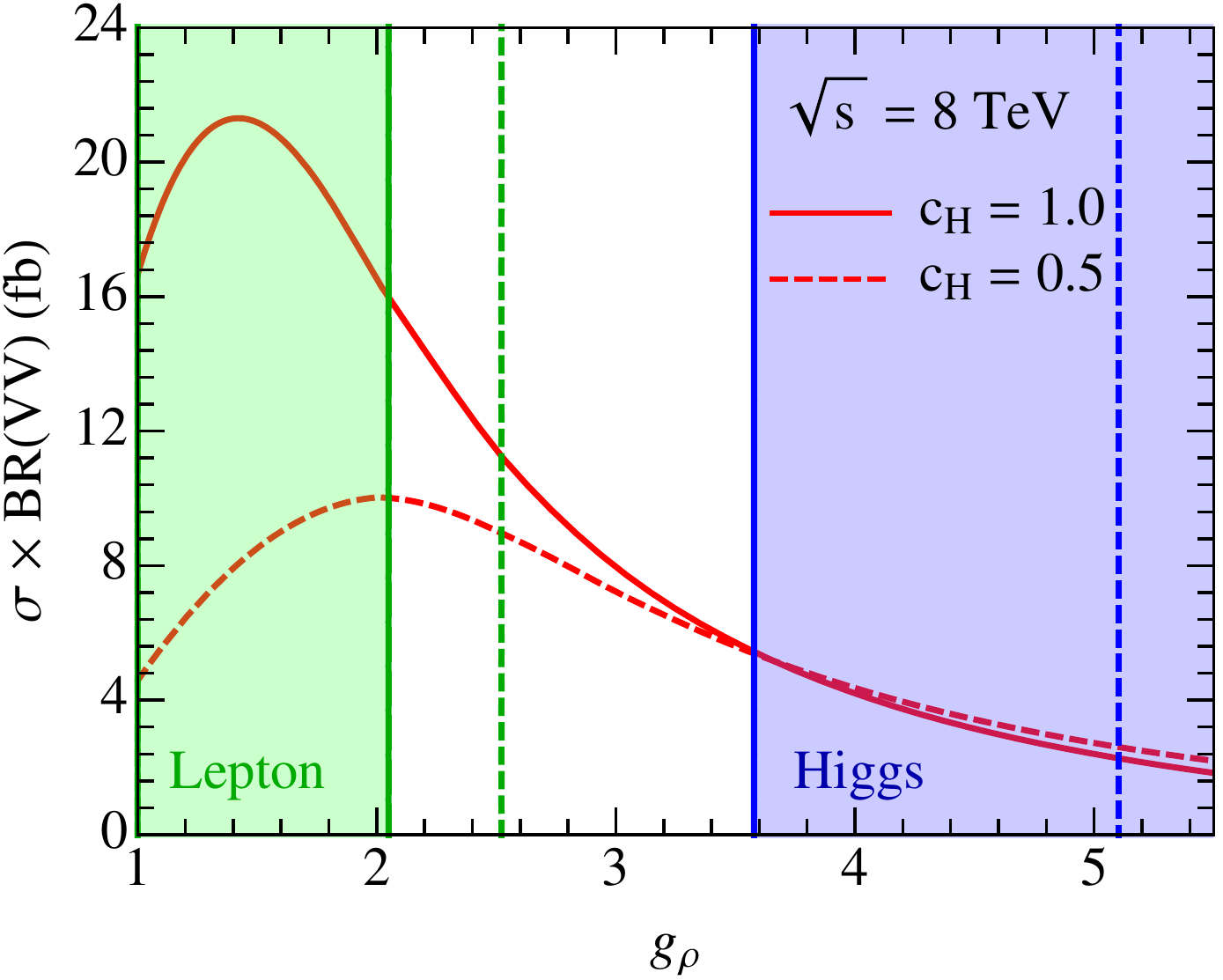}
\caption{``Elementary fermions.''  Diboson rate for $m_\rho =$ 2 TeV with elementary fermions for $c_H = 1$ (solid) and $c_H = 1/2$ (dashed).  Bounds from dilepton searches are shown (green) as are bounds from modifications to Higgs couplings (blue).  As emphasized in the text, this should only be considered as a toy example since it corresponds to the limit of massless fermions.}
\label{fig:sigmaBR-elemFerm}
\end{center}
\end{figure}

We start with the baseline case of no fermion compositeness where the fermion couplings come universally from vector mixing.  As presented in~\cite{Thamm:2015csa}, one finds a sizable diboson rate that passes direct constraints for $2 \lesssim g_\rho \lesssim 3.5$.  We show this in Fig.~\ref{fig:sigmaBR-elemFerm} for $c_H = 1$.  One additionally sees that choosing instead $c_H = 1/2$, as in the two-site model, allows for the range of $2.5 \lesssim g_\rho \lesssim 5$, albeit with a lower overall rate.

Given this as a benchmark, there are two relevant questions brought up by fermion partial compositeness.  The first is whether including fermion compositeness can still accommodate the diboson excess in reasonable regions of parameter space.  This is crucial because full composite Higgs models require fermion mixing for fermion masses.  The second is whether including fermion compositeness opens up new parameter space for smaller couplings $g_\rho \lesssim 2$, or allows for larger rates at larger couplings $g_\rho \simeq 4-5$.

\subsubsection*{Composite top} 

\begin{figure} 
\begin{center}
\includegraphics[width=0.45\textwidth]{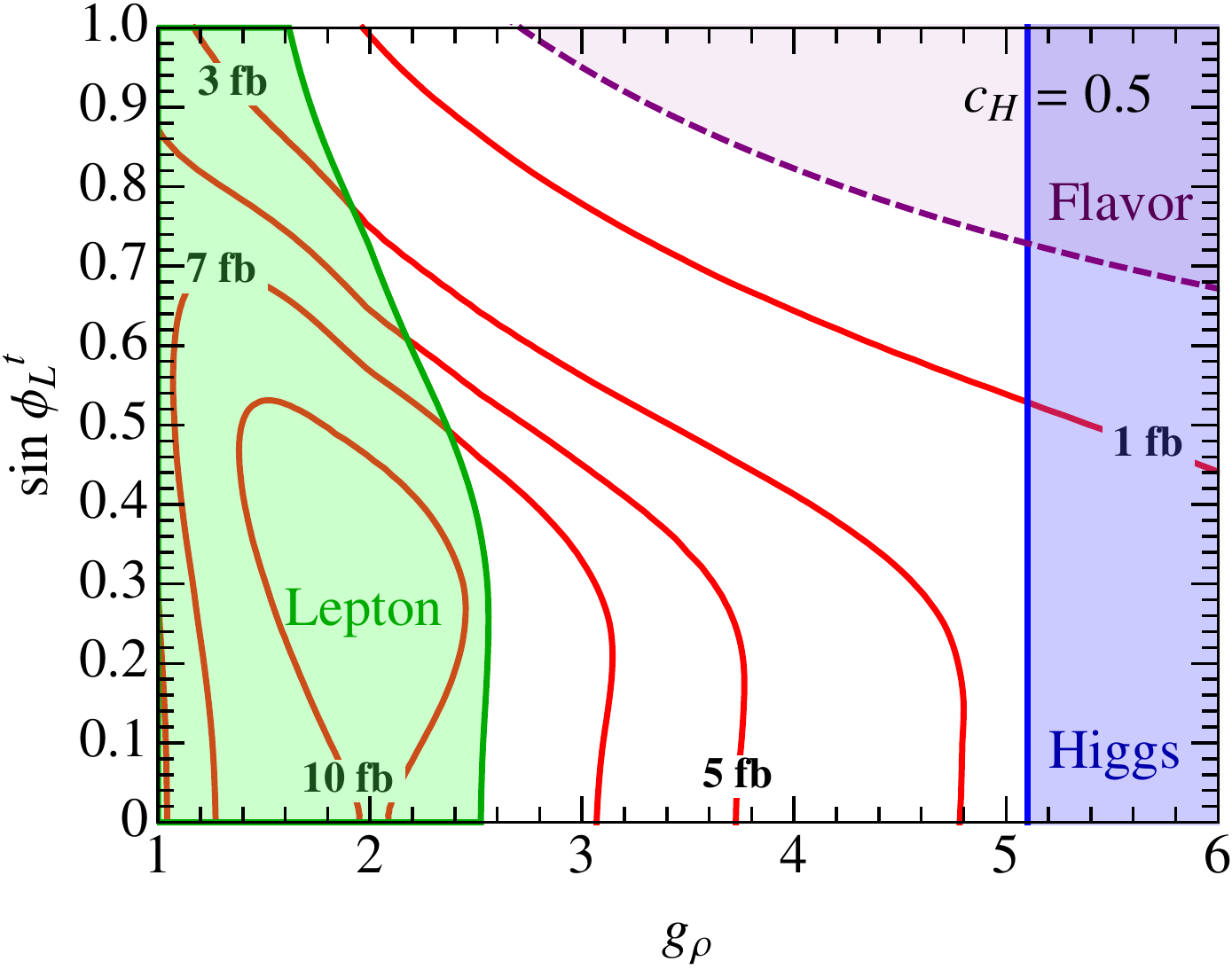} \quad\quad
\includegraphics[width=0.45\textwidth]{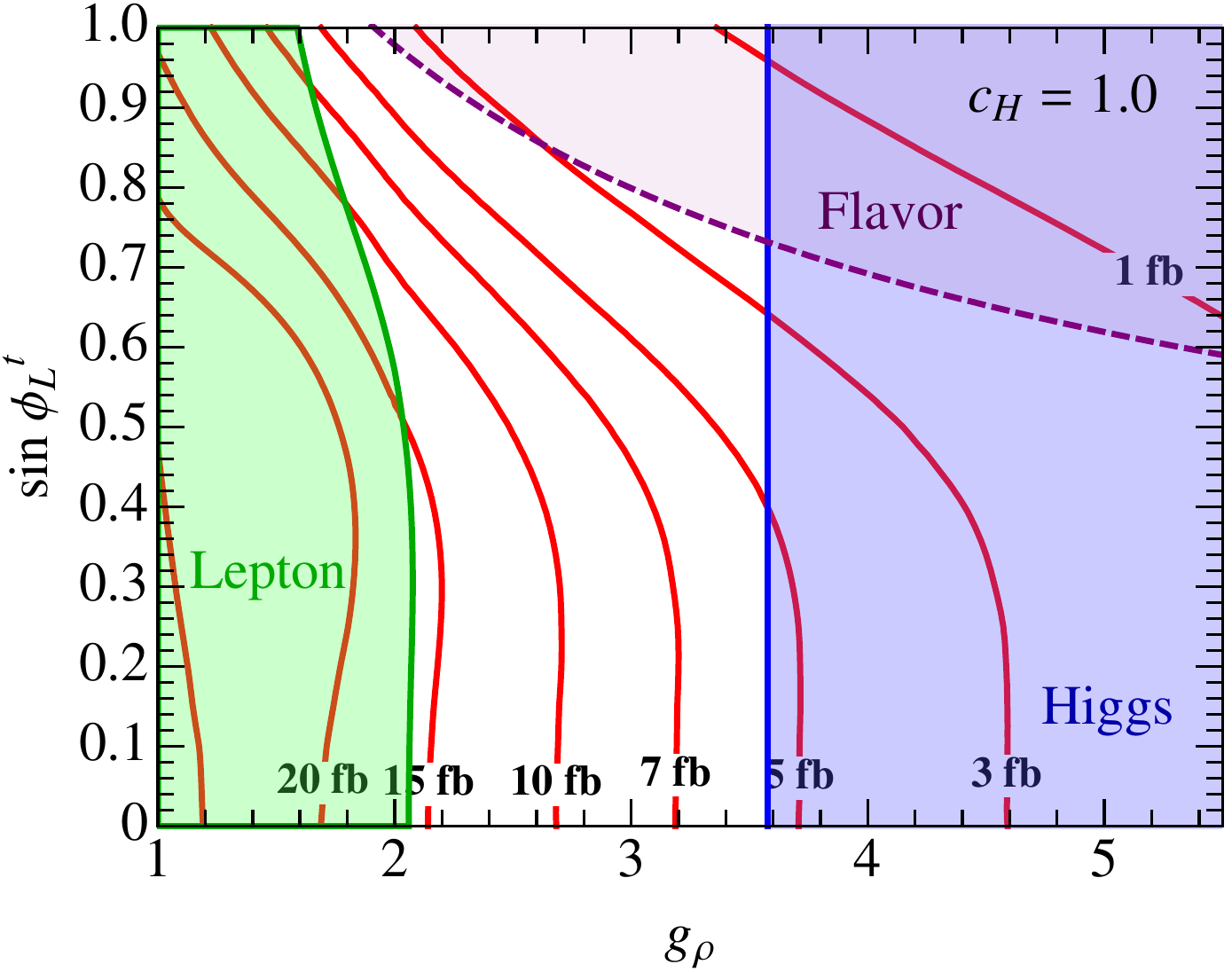}
\caption{``Composite top.''  Diboson rate contours for $m_\rho =$ 2 TeV and $c_H = 1/2$ (left) and $c_H = 1$ (right).  The $y$-axis varies the degree of left compositeness of the $(t_L, b_L)$ multiplet.  The dashed line corresponds to the flavor bound in Eq.~\eqref{eq:sLt-flavor}.}
\label{fig:grhoSinL-compTop}
\end{center}
\end{figure}

Allowing for the top to be composite but keeping the other quarks as elementary yields the same production cross-section as the elementary fermions scenario.  The difference in diboson rates is only due to a diluted branching ratio to dibosons because of the larger coupling to tops, which gets an additional contribution proportional to $s_{L,t}$ (see Table~\ref{tab:couplings}).  The results are shown in Fig.~\ref{fig:grhoSinL-compTop}.  Notice that the $x$-axis corresponds to the elementary fermion scenario.

Including a substantial mixing of the left handed top, one would expect a smaller diboson rate, which is the case for $s_{L,t} \gtrsim 0.5$.  Below these values there are regions where the diboson rates increase slightly.  This is due to cancellations in the couplings to the top which only occur for small $g_\rho$.

As before we also show constraints on the model that originate from other observables, namely dilepton searches and Higgs coupling measurements.  Additionally, here we note that for large values of $s_{L,t}$ bounds from flavor physics are expected.  An exclusion is drawn assuming a U(2) flavor symmetry in the left handed mixings.

As a final remark, we note that scanning over $s_{L,t}$ also scans over different regions of theory space in the sense that larger values of $s_{L,t}$ are natural in standard composite Higgs models while smaller values of $s_{L,t}$ are naturally obtained in composite twin Higgs models.

\subsubsection*{Composite quarks} 

\begin{figure}
\begin{center}
\includegraphics[width=0.45\textwidth]{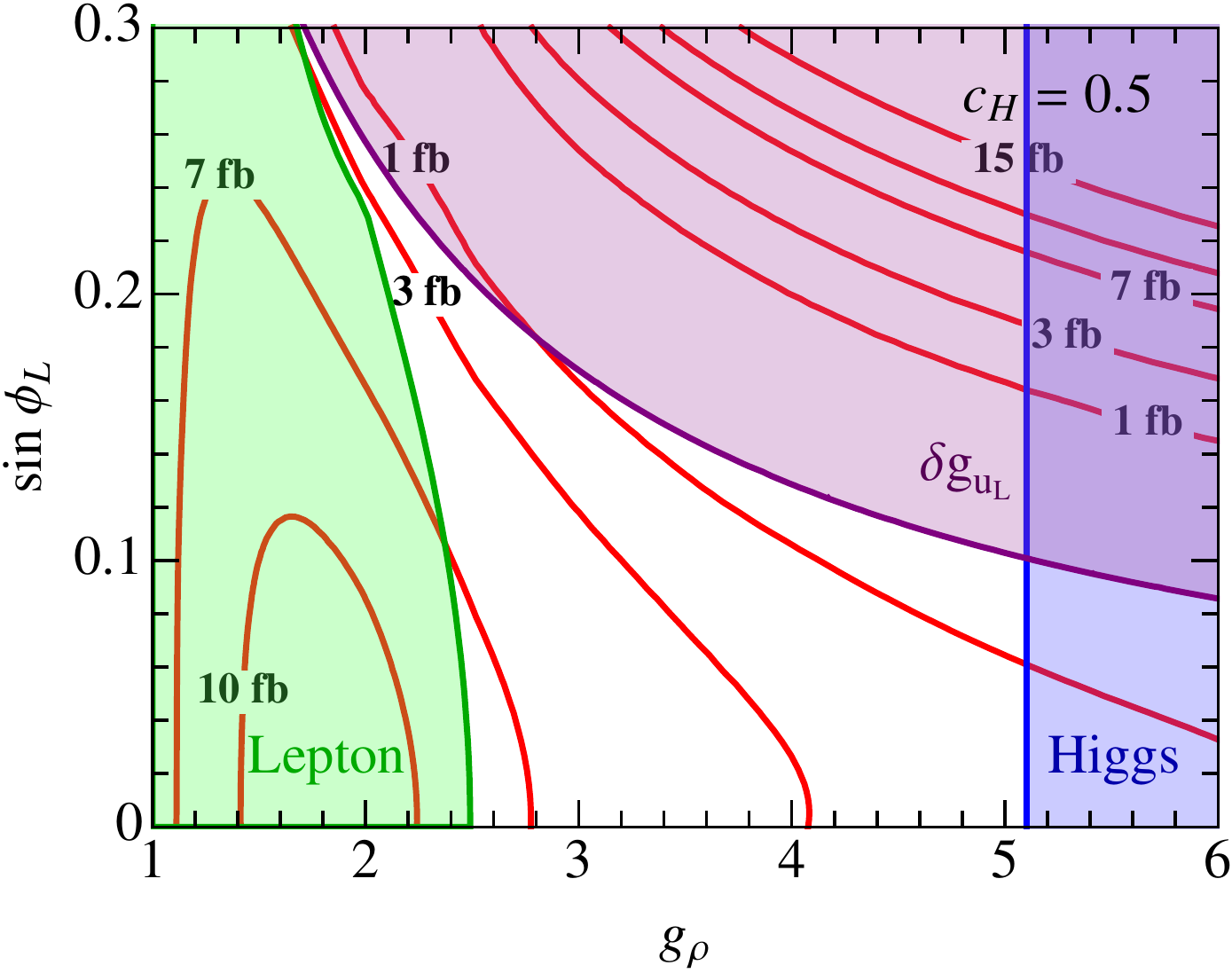} \quad\quad
\includegraphics[width=0.45\textwidth]{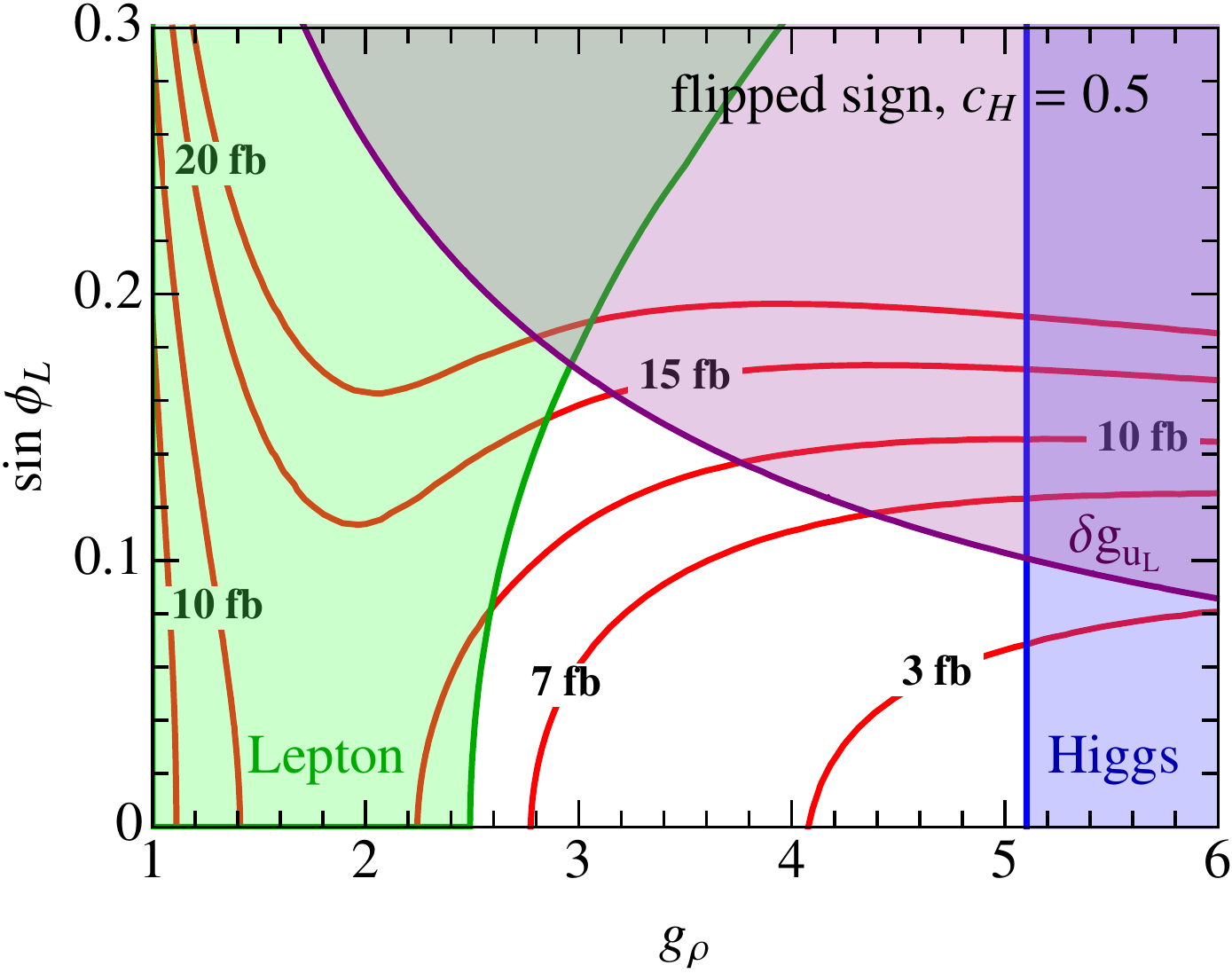}
\caption{``Composite quarks (standard composite Higgs).'' Diboson rate contours for $m_\rho =$ 2 TeV and $c_H = 1/2$ with $a_L=1$ (left) and $a_L= - 1$ (right).  The $y$-axis varies the degree of left compositeness of the $(u_L, d_L)$  and $(c_L, s_L)$ multiplets.  The compositeness of the $(t_L, b_L)$ multiplet is fixed at $\sin\phi_L^t = 0.4$.}
\label{fig:grhoSinL-compQuark}
\end{center}
\end{figure}

\begin{figure}
\begin{center}
\includegraphics[width=0.45\textwidth]{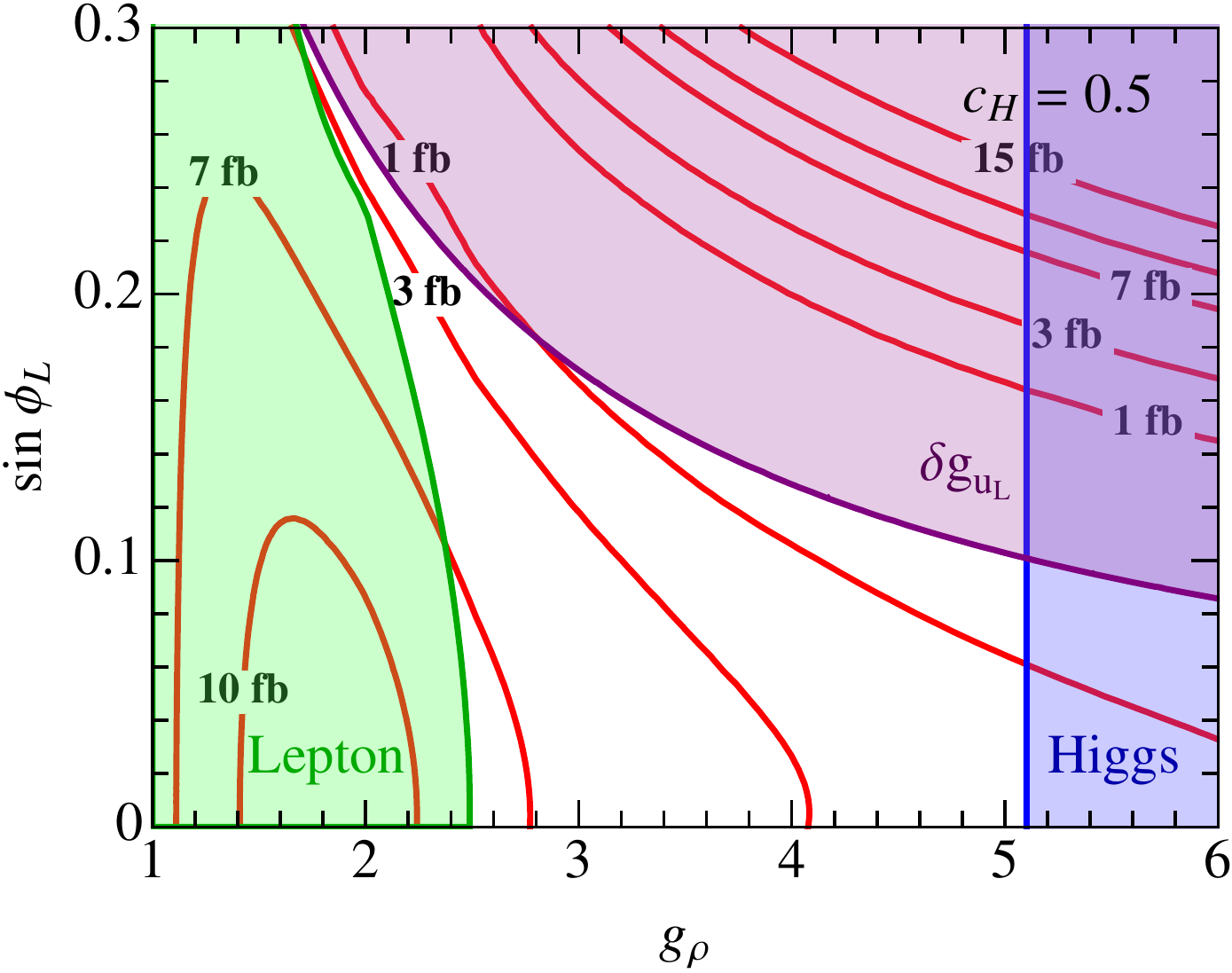} \quad\quad
\includegraphics[width=0.45\textwidth]{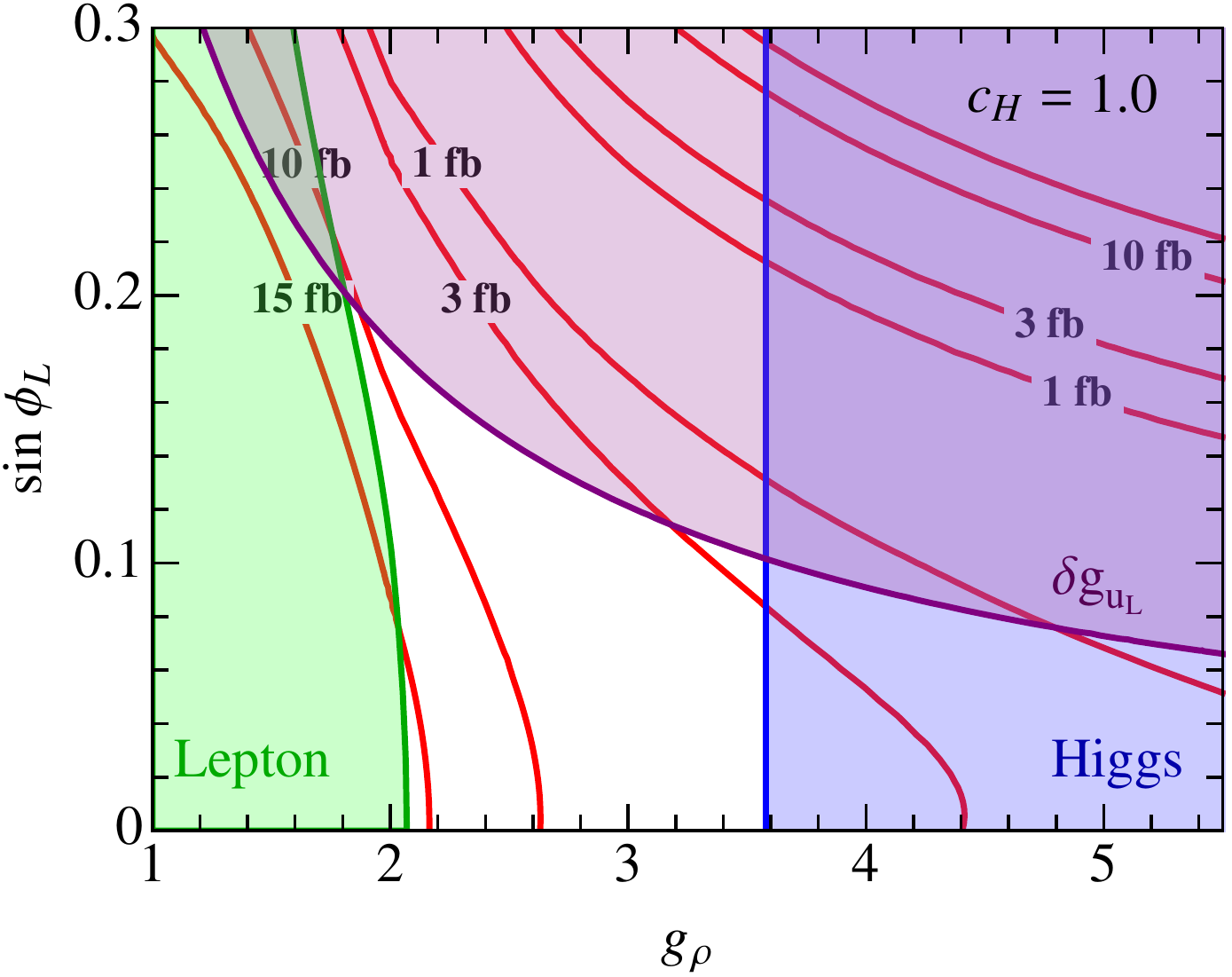}
\caption{``Composite quarks (composite twin Higgs).'' Diboson rate contours for $m_\rho =$ 2 TeV and $c_H = 1/2$ (left) and $c_H = 1$ (right).  The $y$-axis varies the degree of left compositeness of the $(u_L, d_L)$  and $(c_L, s_L)$ multiplets.  The compositeness of the $(t_L, b_L)$ multiplet is fixed at $\sin\phi_L^t = 0.1$.}
\label{fig:grhoSinL-compQuarkTwin}
\end{center}
\end{figure}

The results for the composite quarks scenario are presented in Figs.~\ref{fig:grhoSinL-compQuark} and~\ref{fig:grhoSinL-compQuarkTwin}, where the benchmarks of $s_{L,t} = 0.4$ and $s_{L,t} = 0.1$ have been used.

In these plots the parameter space scanned is the left handed mixing of the lightest two generations of quark doublets set to a common value of $s_{L,q}$.  The top left compositeness is set, as mentioned, by Eq.~\eqref{eq:benchmarks-compQuark}.  Unlike including only top compositeness, now changing the coupling of the light quarks to the vector resonance changes the production rate of the vector.  The effect is to decrease the rates because of the relative minus sign in Table~\ref{tab:couplings}.

Moving away from the two-site model, there is a qualitatively different behavior if we consider $a_L = -1$ shown in the right panel of Fig.~\ref{fig:grhoSinL-compQuark}.  For this case there is no partial cancellation and the rate increases as one increases the quark compositeness.  This is likely preferred for the diboson signal.

In the present picture several other constraints are present.  Besides the usual bounds from dilepton searches and Higgs couplings, we also have bounds from non-universal corrections to precision measurements of the left handed current of the $Z$ and $W$ bosons.  This arises from the fact that as the left mixing of the quarks is increased, there is a larger departure from universality, resulting in larger distortions of the couplings to the $Z$ and $W$ bosons.

\subsubsection*{Role of the $SU(2)_R$ triplet} 

Thus far we have only discussed the spin-1 triplet of SU(2)$_L$.  However, due to the SO(4) symmetry of the strong sector we expect the spin-1 multiplet of SU(2)$_R$ to also play a role.  The SU(2)$_R$ states are almost mass degenerate with the triplet states, with a splitting suppressed by hypercharge.  The couplings to standard model fermions are also determined by hypercharge as shown in Table~\ref{tab:couplings}.  The $\rho_B$ couplings to fermions are suppressed by hypercharge, while the $\rho_C$ does not couple to standard model fermions (before electroweak symmetry breaking) because there are no charged SU(2)$_R$ gauge bosons with which to mix.  Given the hypercharge suppression we can estimate the contribution of $\rho_B$ to the overall rate as $(g'^2 / g^2)^2 \simeq t_w^4 \simeq 0.08$ (for $s_{L,q}\ll 1$).  Thus the $\rho_B$ will increase the rate by roughly 3\% (since its contribution is roughly 10\% of the size of the neutral component $\rho^0$).  In the case of large $s_{L,q}$ (see Table~\ref{tab:couplings}) the rates of $\rho_B$ and $\rho^0$ will become comparable; we avoid very large values of $s_{L,q}$ (for the lighter quarks) in our discussion since these are bounded by precision measurements.

\subsection{Direct constraints}\label{sec:8tevdirect}

During the first run of the LHC, several searches were performed for $W'$ bosons and KK gravitons that can be recast as limits on our model, and have been done so in the figures of the previous section.  In Table~\ref{tab:direct} we report the leading exclusion bounds for a 2 TeV resonance, which correspond to the direct bounds shown in our figures.

\begin{table} [h]
\renewcommand{\arraystretch}{1.2}
\begin{center}
\begin{tabular}{c||rr|rr}
final state & \multicolumn{2}{c|}{ATLAS} & \multicolumn{2}{c}{CMS} \\ \hline \hline
$\ell^+ \ell^-$            & $0.2$ fb & \cite{Aad:2014cka}    & $0.25$ fb & \cite{Khachatryan:2014fba}\\
$\ell^\pm \slashed{E}_T$ & $0.9$ fb & \cite{ATLAS:2014wra}  &  $0.4$ fb & \cite{Khachatryan:2014tva}\\
$t\bar{b}$                 & $120$ fb & \cite{Aad:2014xea}    &  $100$ fb & \cite{Chatrchyan:2014koa}\\
$t\bar{t}$                 &  $50$ fb & \cite{Aad:2015fna}    &   $20$ fb & \cite{Khachatryan:2015sma}\\
$jj$                       & $130$ fb & \cite{Aad:2014aqa}    &  $100$ fb & \cite{Khachatryan:2015sja}
\end{tabular}
\caption{Upper limits (at 95\% CL) on the cross-section $\sigma \times \br$ of a 2 TeV vector decaying to various final states.}
\label{tab:direct}
\end{center}
\end{table}

As we see from the data, the most stringent bound comes from the dilepton channel, closely followed by the single lepton with missing energy channel.  Due to the difference in branching ratios, these two channels happen to constrain parameter space almost identically, so we show only the dilepton bound on the figures for simplicity.  From Eq.~\eqref{eq:brs} we see that to evade the lepton constraints it is sufficient to have $g_\rho \gtrsim 2.5$ for $c_H \simeq 1$.  This constraint becomes slightly weaker as the quarks become more composite because the production rate decreases, as can be seen in Figs.~\ref{fig:grhoSinL-compTop},~\ref{fig:grhoSinL-compQuark}, and~\ref{fig:grhoSinL-compQuarkTwin}.

Like the pair of dilepton and single lepton bounds, the $t\bar{t}$ and $t\bar{b}$ bounds constrain parameter space in the same way.  They are not strong enough to constrain any of the plotted parameter space, but they rule out the composite quark parameter space for large $g_\rho$ and large $s_{L,q}$.  These constraints are always weaker than those from coupling distortions, which are discussed below.

Dijet searches are also not constraining in our plotted parameter space.  These constraints can be meaningful if the dijet branching ratio is very large, like for small $g_\rho$ and large $s_{L,q}$, or if the production rate is very large, like for large $g_\rho$ and large $s_{L,q}$.  Again, coupling distortions are always more constraining.

\subsection{Indirect constraints}\label{sec:8tevindirect}

In this section we discuss the most relevant bounds including Higgs couplings, electroweak parameters, distortions of $W$ and $Z$ couplings, flavor, and searches for compositeness.

\subsubsection*{Higgs couplings} 

Composite Higgs models predict deviations of the Higgs to standard model particles.  While the deviation of Higgs couplings to fermions is dependent on the embeddings of the fermions, the couplings to vectors comes universally from the choice of coset.  In both the minimal composite Higgs and composite twin Higgs models the couplings are predicted to be
\begin{equation}
  c_{hVV} = \sqrt{1 - \frac{v^2}{f^2}} \; c_{hVV}^{\rm SM},
\end{equation}
where $c_{hVV}^{\rm SM}$ is the coupling predicted in the standard model alone.  Current measurements constrain the deviation to be $\lesssim 10\%$, leading to a bound on $f$ of~\cite{ATLAS:2015bea,Khachatryan:2014jba}
\begin{equation}
  f > 550~\mathrm{GeV},
\end{equation}
that appears in all the plots of this section.  Shown in the figures is actually a bound on $g_\rho$ given that we have imposed the relation of Eq.~\eqref{eq:ch-definition}. The high luminosity run of the LHC is expected to increase the bound on $f$ to 800 GeV~\cite{Thamm:2015zwa}.

\subsubsection*{Electroweak parameters} 

Integrating out the composite spin-1 resonances generates a tree-level contribution to the $S$ parameter of the size $\sim 4\pi v^2/m_\rho^2$.  Given the measured value of $S=0.05\pm 0.11$~\cite{Baak:2014ora}, taken alone this bound dictates that $m_\rho \gtrsim 2$ TeV.  In composite Higgs models $S$ and $T$ also receive 1-loop corrections that are sizable and proportional to $c_{S,T}/(16\pi) (v^2/f^2)\log(m_\rho/m_h)$ where $c_{S,T}$ is a calculable coefficient~\cite{Barbieri:2007bh}.  Taking into account the high correlation between $S$ and $T$, the bounds on $m_\rho$ and $f$ are strengthened to the multi-TeV and TeV regions, respectively.  It is however possible to have UV corrections that relax those bounds to $m_\rho \simeq$ 2 TeV and $f \simeq 600$ GeV \cite{Contino:2015mha} (note the latter is comparable to the bound on $f$ from Higgs couplings).

In the model with partially composite fermions, UV corrections can relax the bounds thanks to sizable contributions to the $T$ parameter.  In the two-site model used in this paper, we expect (see \cite{Panico:2015jxa} for a review of possible contributions) a positive correction $\delta T \sim N_c y_t^2/(16\pi^2) (s_{L,t}^2/s_{R,t}^2) (v^2/f^2)$,
that might relax the bound.

While a composite spin-1 resonance of 2 TeV is at the edge of what is allowed by precision tests, possible UV contributions make it difficult to say this definitively.  We believe that a more robust bound, free from many incalculable effects from the strong sector, will be provided by the forthcoming direct exploration at the 13 TeV LHC.

\subsubsection*{Distortions of $W$ and $Z$ couplings} 

Electroweak symmetry breaking induces non-universal corrections to the couplings between $W/Z$'s and standard model fermions.  We write the interaction between the $Z$ and a quark $q$ as
\begin{equation}
\mathcal{L} = 
\frac{g}{\cos\theta_w} Z_\mu \bar{q} \gamma^\mu [ (g^{\rm SM}_{q_L} + \delta g_{q_L}) P_L + (g^{\rm SM}_{q_R} + \delta g_{q_R}) P_R ] q,
\end{equation}
where the standard model couplings are
\begin{equation}
  g^{\rm SM}_{q_L} = T^3_L - Q \sin^2\theta_w,
  \quad\quad\quad\quad\quad
  g^{\rm SM}_{q_R} = - Q \sin^2\theta_w.
\end{equation}
As seen in Table~\ref{tab:couplings} left handed quarks couple the strongest to vector resonances which means the left handed couplings give rise to the tightest constraints on quark compositeness.  Corrections to $\delta g_{q_L}$ and $\delta g_{q_R}$ are constrained by measurements of $R_h$, $R_b$, and the unitarity (of the first row) of the CKM matrix.

Quark compositeness induces a correction of
\begin{equation}
  \delta g_{q_L} = c_{q_L} s_{L,q}^2 \frac{v^2}{f^2},
\end{equation}
with a correction of the same form for right handed quarks and for couplings to $W$'s.  The size of the coefficients $c_{q_L}$ is model dependent, but if the composite sector and mixings respect an approximate left-right symmetry some of these corrections can be highly suppressed~\cite{Agashe:2006at}.

The two-site model we use largely respects the left-right symmetry.  In particular, at leading order in the mixings, we have $c_{u_R} = c_{d_L} = c_{d_R} = 0$.  Notice that the coupling $Z b \bar{b}$ is protected, which strongly relaxes the bound from $R_b$, which would otherwise be very constraining given that $b_L$ has the same mixing as the $t_L$.

The above protection is not at work for the left handed up-type quarks.  If the light up-type quarks have a large mixing, the bounds from $R_h$ and CKM unitarity are still present (for example see~\cite{Redi:2011zi}).  At leading order they are correlated, both being proportional to $\delta g_{u_L}$.  Following~\cite{Matsedonskyi:2014iha}, we consider the $2\sigma$ bound
\begin{equation}
\delta g_{u_L} = \frac{1}{4} \frac{v^2}{f^2} s_{L,u}^2 < 0.5 \times 10^{-3}.
\end{equation}
In practice, one can achieve a smaller coefficient ({\it i.e.} a weaker bound) in front of the coupling modification by changing the parameters of the fermionic contribution.

\subsubsection*{Flavor bounds} 

In the figures describing the composite top scenario we have shown constraints that bound the left mixing of the top.  The origin of this bound can be understood from the fact that we always expect to generate four-fermion operators of the form \cite{Barbieri:2012tu,Matsedonskyi:2014iha}\footnote{At the least the \textbf{6} of vectors can generate them.  The fermion dependent part of the right diagram of Fig.~\ref{fig:rhoToVectors} can generate the effective interaction.  The non-trivial flavor structure arises in the quark physical mass basis.}
\begin{equation}\label{eq:flavor}
\mathcal{L} \supset c'_4 (V_{3i}^* V_{3j})^2 C_{ij}^2 \frac{s_{L,t}^4}{f^2} (\bar{d}_L^i \gamma^\mu d_L^j)(\bar{d}_L^i \gamma_\mu d_L^j)
\end{equation}
where $d$ are standard model down-type quarks of the $i^{\rm th}$ generation, $V_{ij}$ are elements of the CKM matrix, and $C_{ij}$ is a matrix in flavor space that depends on the flavor structure. Given that we assumed an underlying U(2)$^2$ flavor symmetry in the left handed mixings, the strongest bounds come from $\Delta B_s = 2$ observables \cite{Barbieri:2012tu} ($i=2$, $j=3$) which imply
\begin{equation}\label{eq:sLt-flavor}
s_{L,t} \lesssim 0.95  \left( \frac{3}{g_\rho}\right)^{1/2}
\left(\frac{0.5}{c_H}\right)^{1/4}
\left(\frac{0.2}{C_{23}}\right)^{1/2}
\left(\frac{1}{c'_4}\right)^{1/4}
\end{equation}
where we have fixed the relation between $f$ and $m_\rho=$ 2 TeV.  In U(2)$^2$ models $C_{23}$ is a free parameter, but we take it as $C_{23} = 0.2$ (and use $c'_4=1$).

A careful assessment of the flavor bounds is beyond the scope of this work. However, we want to emphasize that in some regions of the parameter space of the composite top scenario, the exclusion corresponding to Eq.~\eqref{eq:sLt-flavor}  can be important (see Fig.~\ref{fig:grhoSinL-compTop}).  Other flavor realizations can lead to different contributions to $C_{ij}$, but typically one cannot make $C_{ij}$ much smaller than one.  In the case of composite quarks the benchmark values of $s_{L,t}$ of Eq.~\eqref{eq:benchmarks-compQuark} satisfy the above bounds in all the parameter space of Figs.~\ref{fig:grhoSinL-compQuark} and \ref{fig:grhoSinL-compQuarkTwin}.

\subsubsection*{Compositeness constraints} 

In the composite quarks scenario an additional constraint comes from searches for quark compositeness~\cite{Khachatryan:2014cja,Aad:2015eha}.  The experimental results are usually presented as bounds on four-fermion operators $(2\pi/\Lambda^2)(\bar{q} \gamma^\mu q)(\bar{q} \gamma_\mu q)$, with bounds of the order $\Lambda \gtrsim 10$ TeV.  In our scenario, analogously to Eq.~\eqref{eq:flavor}, the strong sector produces effects proportional to
\begin{equation}
\mathcal{L} \supset c_4 \frac{s_{L,q}^4}{f^2} (\bar{q}_L \Gamma^\mu q_L)(\bar{q}_L \Gamma_\mu q_L)
\end{equation}
where $\Gamma^\mu$ represents the possible Lorentz and color structure, and $c_4$ is an $\mathcal{O}(1)$ coefficient that we take to be 1 (see~\cite{Domenech:2012ai} for a careful computation of it).  Here we consider only the left handed quarks.  This implies a bound on $s_{L,q}$ (taken to be the same for the lightest two generations) of
\begin{equation}
s_{L,q} \lesssim 0.5
\left(\frac{3}{g_\rho}\right)^{1/2} \left(\frac{0.5}{c_H}\right)^{1/4}
\left(\frac{1}{c_4}\right)^{1/4}.
\end{equation}
This bound is milder than the precision constraints.

%
%

\section{Projections for 13 TeV LHC data}\label{sec:13tev}

\begin{figure}
\begin{center}
\includegraphics[width=0.45\textwidth]{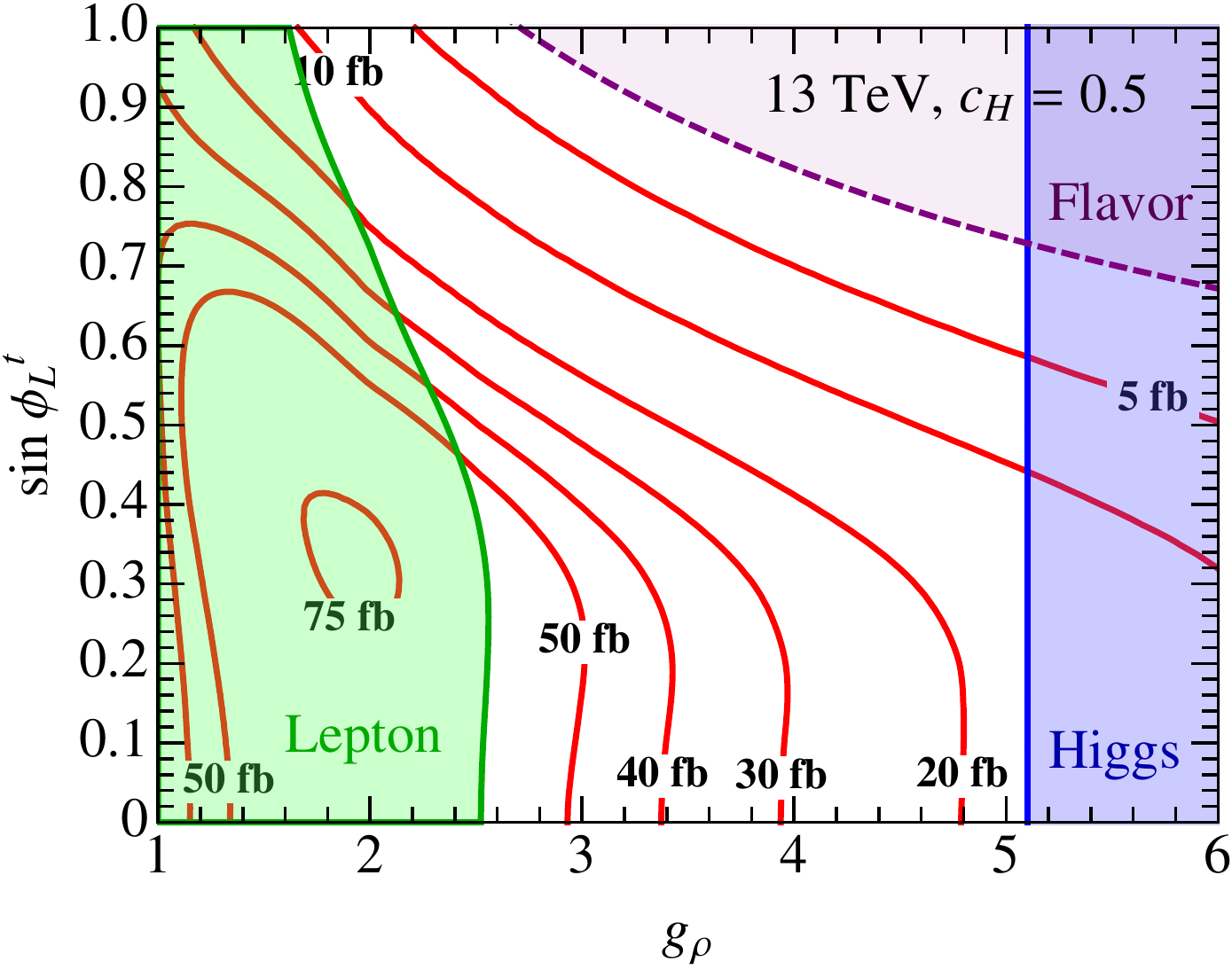} \quad\quad
\includegraphics[width=0.45\textwidth]{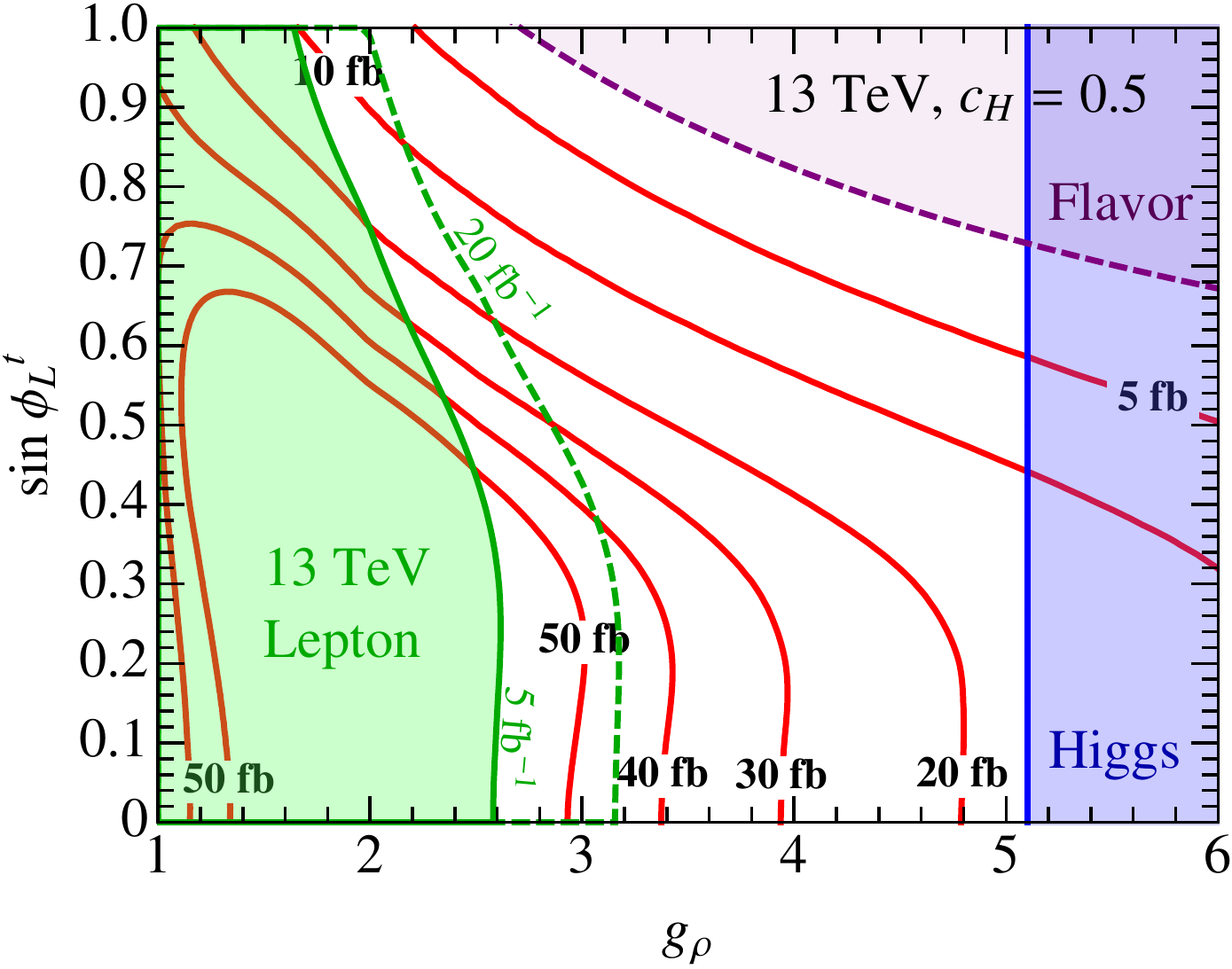}
\caption{``Composite top.''  Diboson rate contours at 13 TeV, using $c_H=1/2$. Left panel: bounds from current 8 TeV lepton searches. Right panel: projected bounds from 13 TeV lepton searches.}
\label{fig:grhoSinL-compTop-13tev}
\end{center}
\end{figure}

In this section, we make some preliminary projections for the prospect of searching for spin-1 composite resonances in 13 TeV LHC data.  If the diboson signal is real, it is interesting to examine how this will be probed at the next run of the LHC.\footnote{See also \cite{Abe:2015uaa}.}   If not, we also present the signal rates for heavier composite resonances.  In this section we focus on our composite top scenario as an example (though recall that this still corresponds to the $s_{L,q} = 0$ limit of the composite quarks scenario).  We start by computing diboson rates at 13 TeV, shown in Fig.~\ref{fig:grhoSinL-compTop-13tev} (left).  Here the exclusion limits correspond to the current experimentally measured limits.

Next, we evaluate the impact of future dilepton searches on either confirming or ruling out the diboson signal in the context of composite Higgs.  The 13 TeV dilepton exclusion limit can be estimated by assuming the exclusion reach is determined by the number of background events in a window close to the resonance mass~\cite{Thamm:2015zwa}. Under this assumption one can use existing exclusion limits and rescale them to different collider energies and luminosities (see also \cite{SalamWeiler}).  We find an excluded dilepton cross-section of $\sigma = 1.2$ fb with 5 fb$^{-1}$ and $\sigma = 0.4$ fb with 20 fb$^{-1}$.

Figure~\ref{fig:grhoSinL-compTop-13tev} shows that the increased exclusion reach of dilepton searches will probe much space of the region that is best suited for the explanation of the diboson excess, \textit{i.e.} $g_\rho \simeq 2-3$ as can be understood by comparing with Fig.~\ref{fig:grhoSinL-compTop}.  Even a few femtobarns of integrated luminosity is likely sufficient for dilepton searches to make meaningful statements about the diboson excess.

The same rescaling procedure can be applied to current LHC $W^\pm Z$ diboson limits (including only the contribution from $\rho^\pm$) and we find an excluded cross-section of $\sigma = 57-66$ fb with 5 fb$^{-1}$ and $\sigma = 24-26$ fb with 20 fb$^{-1}$.  There is a range of values because the background for diboson searches is comprised of $q\bar{q}$, $qg$, and $gg$ initial states (while for dilepton it is just $q\bar{q}$ at leading order).  As is the case at 8 TeV, dilepton searches are constraining for small values of $g_\rho$ but diboson searches will be more sensitive, in general, due to the larger branching ratio.

\begin{figure}
\begin{center}
\includegraphics[width=0.45\textwidth]{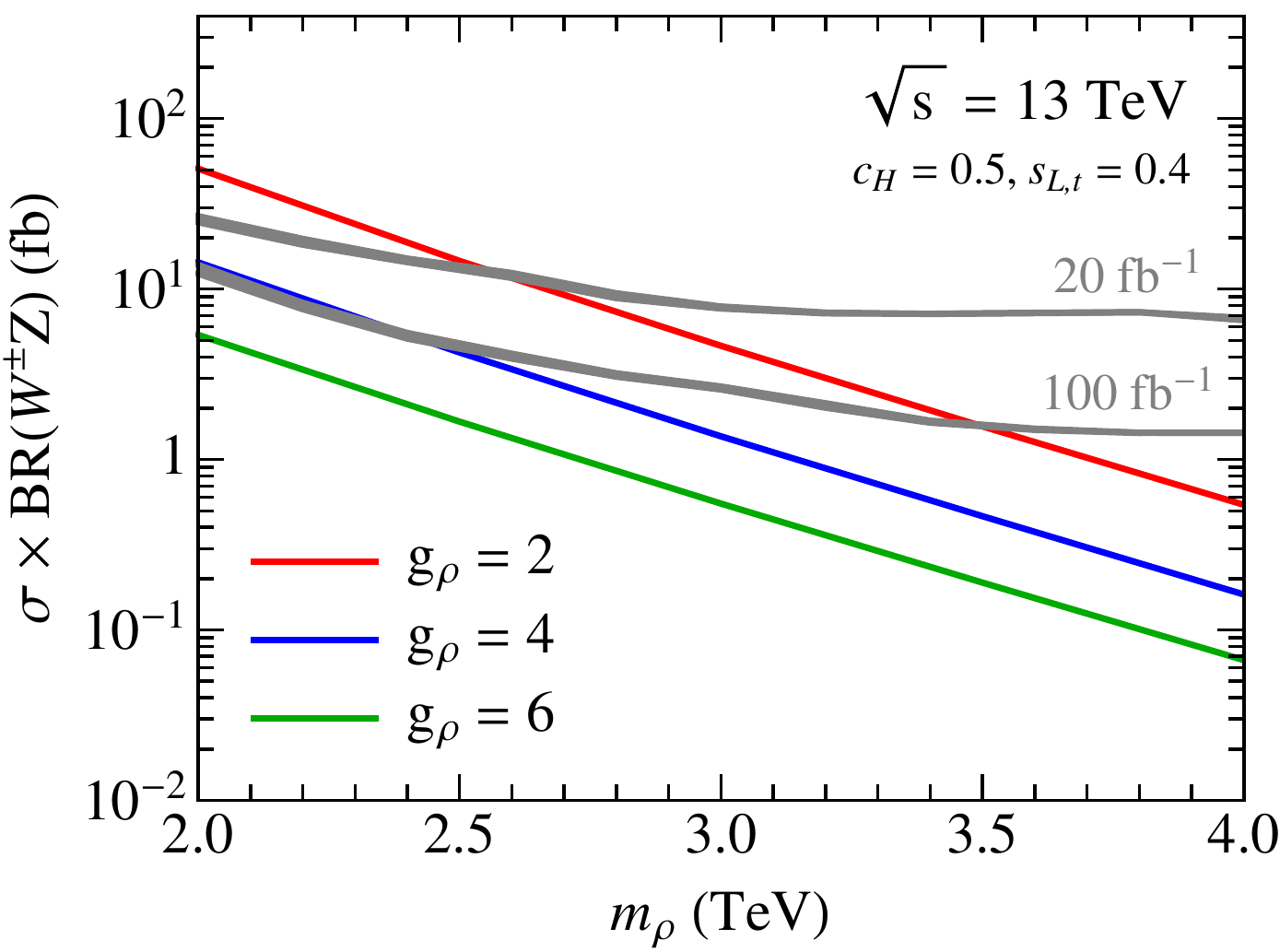} \quad\quad
\includegraphics[width=0.45\textwidth]{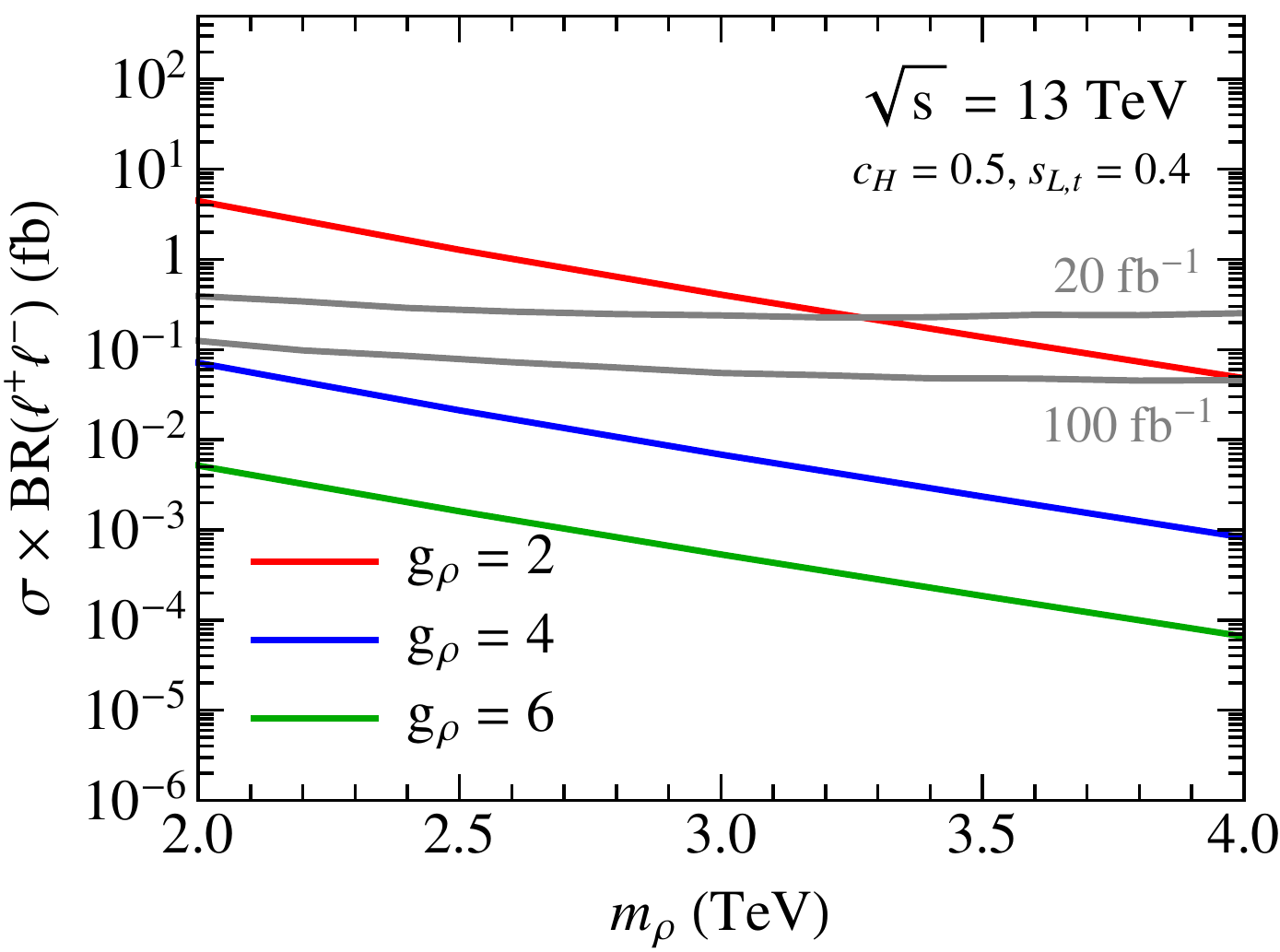}
\caption{``Composite quarks (standard composite Higgs).''  13 TeV cross-sections for diboson (left) and dilepton (right) for $s_{L,t}=0.4$ and $s_{L,q}=0$.  The gray lines show the projected limits with 20~fb$^{-1}$ and 100~fb$^{-1}$ of integrated luminosity.}
\label{fig:massXsec-compQuark-13tev}
\end{center}
\end{figure}

\begin{figure}
\begin{center}
\includegraphics[width=0.45\textwidth]{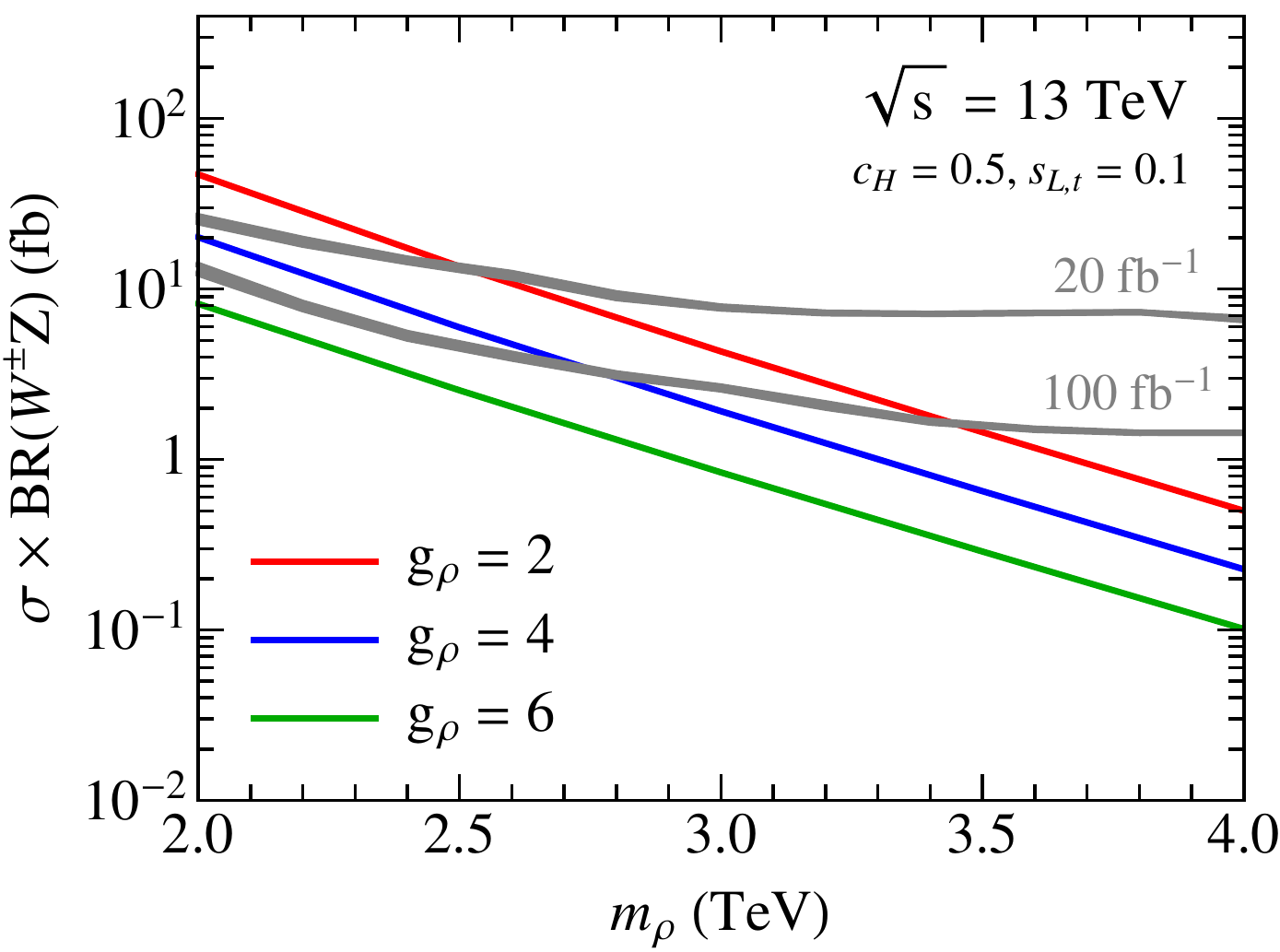} \quad\quad
\includegraphics[width=0.45\textwidth]{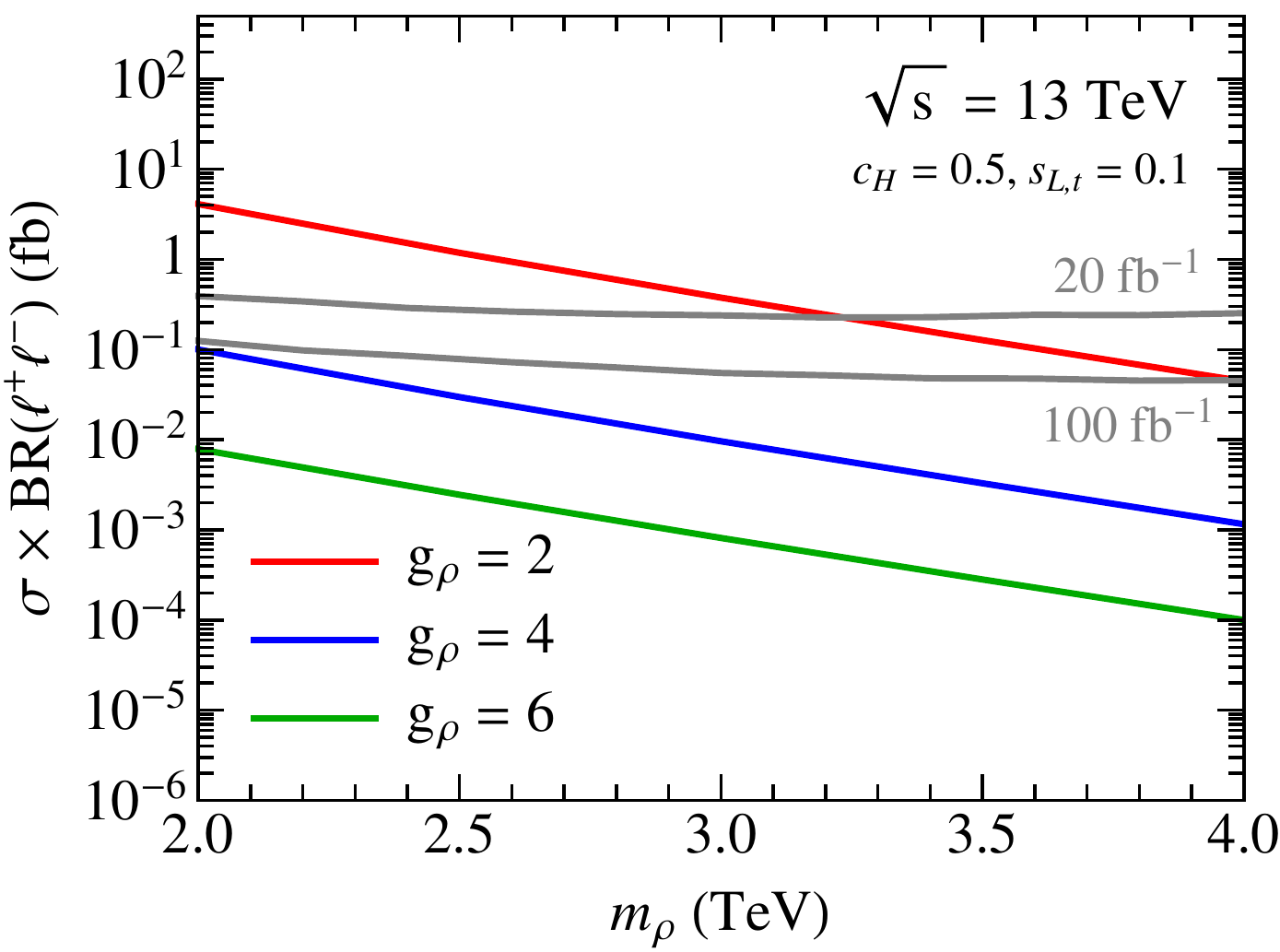}
\caption{``Composite quarks (composite twin Higgs).''  13 TeV cross-sections for diboson (left) and dilepton (right) for $s_{L,t}=0.1$ and $s_{L,q}=0$.  The gray lines show the projected limits with 20~fb$^{-1}$ and 100~fb$^{-1}$ of integrated luminosity.}
\label{fig:massXsec-compQuarkTwin-13tev}
\end{center}
\end{figure}

We also take a closer look at the composite quark scenario in Figs.~\ref{fig:massXsec-compQuark-13tev} and~\ref{fig:massXsec-compQuarkTwin-13tev}.  In these two figures we use the different benchmark values of $s_{L,t} = 0.4$ in Fig.~\ref{fig:massXsec-compQuark-13tev} and $s_{L,t} = 0.1$ in Fig.~\ref{fig:massXsec-compQuarkTwin-13tev}, as representative values of standard composite Higgs and composite twin Higgs.  We scan over the mass of the resonance allowing for values larger than 2 TeV.  In these figures the gray bands correspond to projected limits computed by the rescaling described above.  For the diboson rates the thickness of the band is spanned by varying the background composition between $q\bar{q}$ and $gg$ ($qg$ falls in between them) while for dilepton rates the background is from $q\bar{q}$.


\section{Conclusions}\label{sec:conclusions}

In this paper we have discussed the excess in the ATLAS diboson data in connection with composite Higgs models, where the signal of a multiplet of massive vector bosons can fit the data.  This class of models features a natural enhancement of the coupling between the composite resonances and the longitudinal modes of the $W^\pm$ and $Z$, since, as with the Higgs, they are part of the composite sector. There is also a suppression of the coupling to the standard model fermions. This property makes it plausible that the diboson channel is the leading discovery mode.  In the minimal composite Higgs scenario with a spontaneous breaking of SO(5)/SO(4), we expect a complete multiplet of vector resonances in the \textbf{6} of SO(4), with a mass close to the TeV scale for naturalness considerations.  We have used a concrete two-site model~\cite{Panico:2011pw} that allows us to compute observables and quantitatively describes the above picture.  Where relevant, we have also considered possible deviations from the simplified description of the two-site model. 

\begin{figure}
\begin{center}
\includegraphics[width=0.5\textwidth]{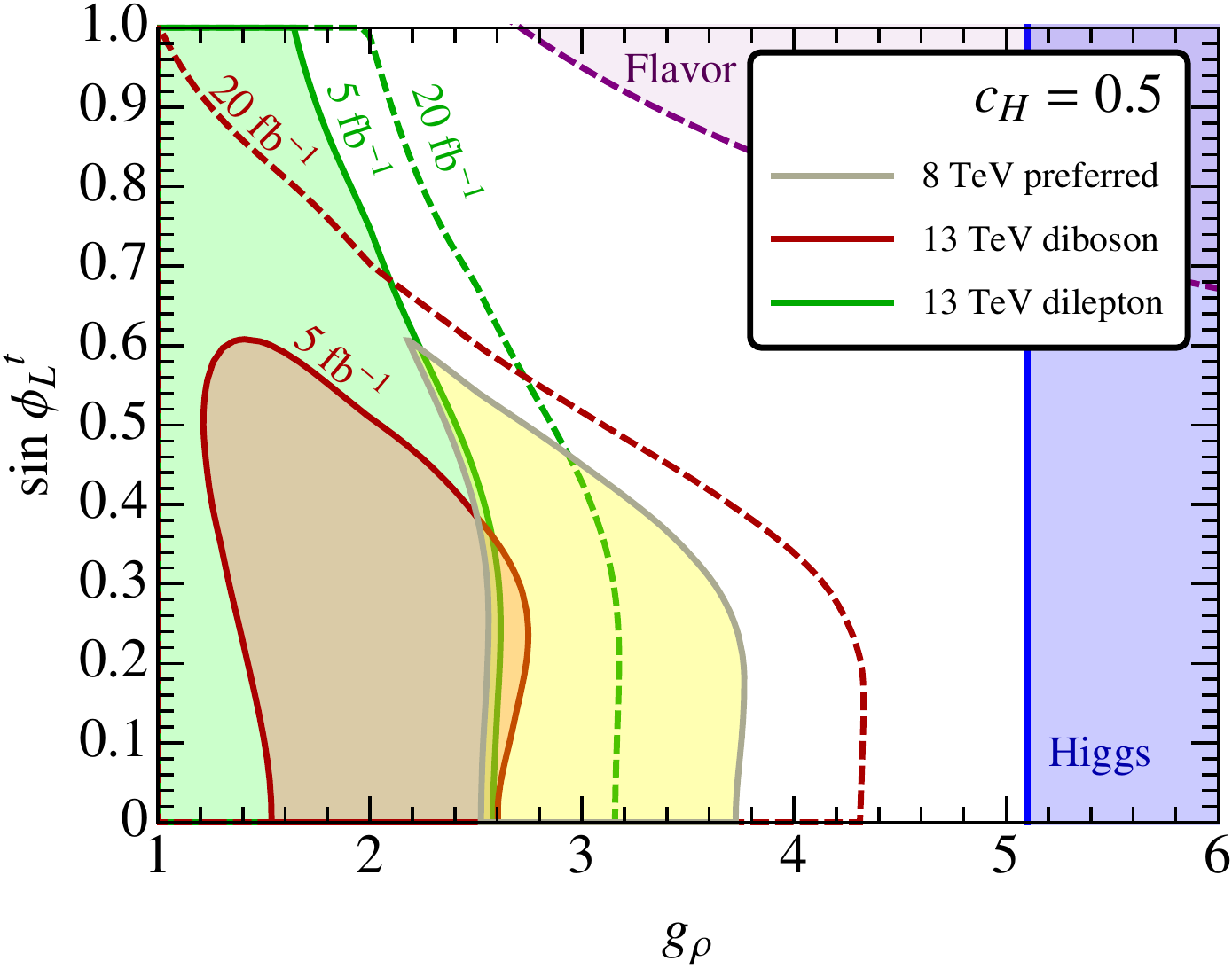}
\caption{``Composite top'' (using $c_H=1/2$).  The yellow region shows the parameter space that describes the 8 TeV diboson excess (see Sec.~\ref{sec:8tev}).  The red and green regions show the projected 95\% CL limits of diboson and dilepton searches, respectively, at 13 TeV (see Sec.~\ref{sec:13tev}).}
\label{fig:grhoSinL-compTop-13tev-lumi}
\end{center}
\end{figure}

This class of composite Higgs  models predicts one SU(2)$_L$ triplet with hypercharge $Y=0$ and three SU(2)$_L$ singlets with hypercharges $Y=0$ and $Y=\pm 1$. The setup itself is not new and has already been studied in depth.  It is useful because it is predictive; the mass of the resonance is related to its coupling strength times the scale $f$ where the extended global symmetry is broken, $m_\rho \sim g_\rho f$.  Since naturalness suggests $f$ be not too much above the weak scale (for the experimental lower bound see the discussion in Sec.~\ref{sec:8tevindirect}), in order to have a mass of $m_\rho \sim 2$ TeV, we need a coupling of $g_\rho \simeq 2-3$.   In most of the scenarios that we have analyzed, the coupling of the composite vectors to standard model fermions is proportional to $g^2/g_\rho$, which is not too suppressed for this range of $g_\rho$.  Indeed, for larger values of $g_\rho$ the production cross-section quickly falls.  It is interesting that the numerics for $g_\rho \simeq 2-3$ can describe the ATLAS diboson excess, while passing all other direct and indirect limits, as we demonstrated in Fig.~\ref{fig:sigmaBR-elemFerm}.

It is necessary to move beyond the (over)simplified case of elementary fermions and consider the realistic case of fermion compositeness.  Indeed, this is a generic feature in composite Higgs models.  A priori,  moving away from the simplified picture of elementary fermions would induce significant changes in the signal rates due to modified couplings. In addition,  fermion compositeness also faces a number of potentially strong indirect constraints which may limit the possible range of signal rates.  In this work we analyzed two cases, one with only the third generation quarks being partially composite and another case, inspired by flavor symmetries, where the lightest two generation of quarks have the same degree of left compositeness, constrained by precision measurements.  Interestingly, there is parameter space in both of these realistic scenarios that can describe the excess.

Independent of the exact degree of quark compositeness, we expect that important complementary information will come from early 13 TeV LHC data.  In all models we found that dilepton and single lepton searches provide a robust bound that will encroach, with as little as 20 fb$^{-1}$ of 13 TeV data, into the preferred parameter space.  Of course, diboson searches are also very important and we have presented projections for the sensitivity of this channel too.  This is an obvious measurement that will need to be made to really understand the nature of this excess.  As an example, in the case of a composite top, we show in Fig.~\ref{fig:grhoSinL-compTop-13tev-lumi} the expected coverage provided by 13 TeV diboson and dilepton searches compared to the parameter space that describes the diboson excess.

In composite Higgs models, the composite resonances also couple strongly to the Higgs boson.  In addition to the $WW$ and $WZ$ channels analyzed in this paper, there are correlated channels like $\rho \to Vh$ with similar rates, providing additional discovery channels.  A detailed analysis of these channels is beyond the scope of this paper.  

The fact that the ATLAS diboson excess can be explained within the framework of natural composite Higgs models is very compelling and certainly warrants further investigation, both from the experimental side and from the theoretical side.

\subsubsection*{Acknowledgements}

AT is supported by an Oehme Fellowship and LTW and ML are supported by DOE grant DE-SC0013642. AT thanks SISSA for hospitality and partial support.

\appendix
\section{Two-site model}\label{app:two-site}

The two-site model starts with an enlarged global symmetry of SO(5)$_1$ $\times$ SO(5)$_2$ / SO(5)$_D$, where SO(5)$_D$ is the diagonal combination.  The coset yields 10 Goldstones (in the adjoint of SO(5)), which can be classified according to an SO(4) subgroup of SO(5)$_D$ as a \textbf{4} which corresponds to the Higgs multiplet and a \textbf{6} which is eaten to yield the massive vector multiplet which we have been discussing.  To get the right fermion gauge numbers there is an additional global U(1)$_X$ that acts on both sites and is used to define hypercharge $Y = T_R^3 + X$.

The Lagrangian, in the elementary-composite basis, is
\begin{equation}\label{eq:ele-comp}
\mathcal{L}^{\rm 2-site} =
 \frac{f^2}{4}\mathrm{Tr}[(D_\mu U)^T D^\mu U]
 + \mathcal{L}^{\rm 2-site}_{\rm vectors}
 + \mathcal{L}^{\rm 2-site}_{\rm fermions},
\end{equation}
where $\mathcal{L}^{\rm 2-site}_{\rm vectors}$ includes the vector kinetic terms and $\mathcal{L}^{\rm 2-site}_{\rm fermions}$ includes the fermion mixing and will be discussed below.  The matrix $U$ contains the Goldstones and is
\begin{equation}
U = \exp\left( i \frac{\sqrt{2}h^i\hat{T}^i}{f} \right),
\quad\quad\quad\quad
i=1,2,3,4,
\end{equation}
where $\hat{T}^i$ are the broken generators of SO(5)$_D$ and $h^i$ are the Goldstones.  The covariant derivative acts on $U$ as
\begin{equation}
D_\mu U= \partial_\mu U - i \hat{g} A_\mu^a T_L^a U -i \hat{g}' B_\mu T_R^3 U + i \hat{g}_\rho U \hat{\rho}_\mu^b T^b.
\end{equation}
The index $a$ runs over $1,2,3$, and $b$ runs over $1,2,3,4,5,6$.

\subsection{Partially composite vectors}\label{sec:pcvectors}

The vector part of the Lagrangian is
\begin{equation}\label{eq:ele-comp-vectors}
 \mathcal{L}^{\rm 2-site}_{\rm vectors}
 =
 -\frac{1}{4} \hat{W}^a_{\mu\nu}{}^2
 -\frac{1}{4} \hat{B}_{\mu\nu}{}^2
 -\frac{1}{4} \hat{\rho}^b_{\mu\nu}{}^2
\end{equation}
where $a$ runs over $1,2,3$ and $b$ runs over $1,2,3,4,5,6$.  The $\hat{W}$, $\hat{B}$, and $\hat{\rho}$ fields are the elementary SU(2)$_L$, elementary U(1)$_Y$, and composite SO(4) fields, respectively.  Their associated gauge couplings are denoted $\hat{g}$, $\hat{g}'$, and $g_\rho$.  These terms are expanded as
\begin{subequations}
\begin{align}
\hat{W}_{\mu\nu}^a     &= \partial_\mu \hat{W}_\nu^a - \partial_\nu \hat{W}_\mu^a + \hat{g} \epsilon^{abc} \hat{W}_\mu^b \hat{W}_\nu^c , \\
\hat{B}_{\mu\nu}       &= \partial_\mu \hat{B}_\nu   - \partial_\nu \hat{B}_\mu , \\
\hat{\rho}_{\mu\nu}^a  &= \partial_\mu \hat{\rho}_\nu^a - \partial_\nu \hat{\rho}_\mu^a + g_\rho f^{abc} \hat{\rho}_\mu^b \hat{\rho}_\nu^c .
\end{align}
\end{subequations}
Here $f^{abc}$ are the structure constants of SO(4).  Since the SU(2)$_L$ and SU(2)$_R$ subgroups of SO(4) commute, the $\hat{\rho}$ field can be split up into an SU(2)$_L$ field $\hat{\rho}_L$ and an SU(2)$_R$ field $\hat{\rho}_R$.

Eq.~\eqref{eq:ele-comp-vectors} is not in a physical basis, but we can rotate to the physical basis before electroweak symmetry breaking by identifying the massless standard model gauge fields by diagonalizing the mass matrix which arises from Eq.~\eqref{eq:ele-comp}.
\begin{equation}
  \mathcal{L} \supset 
  \frac{f^2}{4}( \hat{g} \hat{W}^\mu - g_\rho \hat{\rho}_L^\mu )^2 
  + \frac{f^2}{4}( \hat{g}' \hat{B}^\mu - g_\rho \hat{\rho}_R^{3\mu} )^2 
\end{equation}
The SU(2) piece is diagonalized by the rotation
\begin{equation}\label{eq:rot-2-vector}
\begin{aligned}
  \hat{W}^a      &= c_2 W^a - s_2 \rho_L^a, \\
  \hat{\rho}_L^a &= s_2 W^a + c_2 \rho_L^a,
\end{aligned}
\quad\quad\quad\quad
s_2 \equiv \frac{\hat{g}}{\sqrt{g_\rho^2 + \hat{g}^2}}, \quad
c_2 \equiv \frac{g_\rho}{\sqrt{g_\rho^2 + \hat{g}^2}},
\end{equation}
and similarly for the U(1) piece
\begin{equation}\label{eq:rot-y-vector}
\begin{aligned}
  \hat{B}        &= c_y B - s_y \rho_B, \\
  \hat{\rho}_R^3 &= s_y B + c_y \rho_B,
\end{aligned}
\quad\quad\quad\quad
s_y \equiv \frac{\hat{g}'}{\sqrt{g_\rho^2 + \hat{g}'^2}}, \quad
c_y \equiv \frac{g_\rho}{\sqrt{g_\rho^2 + \hat{g}'^2}}.
\end{equation}
Eqs.~\eqref{eq:rot-2-vector} and~\eqref{eq:rot-y-vector} define the gauge couplings
\begin{subequations}
\begin{align}
  \frac{1}{g^2}  &= \frac{1}{\hat{g}^2} + \frac{1}{g_\rho^2}, \\
  \frac{1}{g'^2} &= \frac{1}{\hat{g}'^2} + \frac{1}{g_\rho^2}.
\end{align}
\end{subequations}
Before electroweak symmetry breaking we now have massless $W^a$ and $B$ fields and massive $\rho$ fields with masses
\begin{equation}\label{eq:spectrum}
\begin{aligned}
\mathbf{3}_0   &: m_{\rho_L}^2 = m_\rho^2 \; (1+ \nicefrac{\hat{g}^2}{g_\rho^2}), \\
\mathbf{1}_0   &: m_{\rho_B}^2 = m_\rho^2 \; (1+ \nicefrac{\hat{g}'^2}{g_\rho^2}), \\
\mathbf{1}_\pm &: m_{\rho_C}^2 = m_\rho^2,
\end{aligned} \quad\quad\quad m_\rho^2 = \frac{g_\rho^2 f^2}{2}.
\end{equation}
explicitly showing our earlier statement that all vectors in the \textbf{6} are approximately mass degenerate.

\subsection{Partially composite fermions}\label{sec:pcfermions}

The fermion part of the Lagrangian is given by
\begin{equation}\label{eq:ele-comp-fermions}
\begin{aligned}
 \mathcal{L}^{\rm 2-site}_{\rm fermions} = \;\;
 & \bar{q}_L i \slashed{D} q_L
 + \bar{u}_R i \slashed{D} u_R
 + \bar{\Psi}_4 (i\slashed{D} - m_4) \Psi_4
 + \bar{\Psi}_1 (i\slashed{D} - m_1) \Psi_1 \\
 & + y_L f \bar{q}_L (U \Psi)
 + y_R f \bar{u}_R (U \Psi) + \mathrm{h.c.}.
\end{aligned}
\end{equation}
The $q_L$ and $u_R$ denote elementary fermion fields while $\Psi_4$ and $\Psi_1$ denote strong sector fields.\footnote{There are also the down-type fields $d_R$ and an associated $\Psi_1$ field, but we omit these for simplicity as their effects are suppressed by $y_b$.}  We consider the \textbf{5} + \textbf{5} model in which $\Psi$ is a \textbf{5}$_{2/3}$ of SO(5) $\times$ U(1)$_X$ which contains $\Psi_4$, a \textbf{4} of SO(4), and $\Psi_1$, a \textbf{1} of SO(4).  For down-type quarks, the embedding is similar except that one needs an $X=-1/3$ multiplet.

The covariant derivatives are defined as
\begin{subequations}
\begin{align}
D_\mu q_L =& \left(\partial_\mu - i \hat{g} \hat{W}_\mu - i \frac{1}{6} \hat{g}' \hat{B}_\mu \right) q_L, \\
D_\mu u_R =& \left(\partial_\mu - i \frac{2}{3} \hat{g}' \hat{B}_\mu \right) u_R, \\
D_\mu \Psi_4 =& \left(\partial_\mu - i \frac{2}{3} \hat{g}' \hat{B}_\mu - i g_\rho \hat{\rho}_\mu \right) \Psi_4, \\
D_\mu \Psi_1 =& \left(\partial_\mu - i \frac{2}{3} \hat{g}' \hat{B}_\mu \right) \Psi_1.
\end{align}
\end{subequations}
The \textbf{4}$_{2/3}$ field $\Psi_4$ decomposes into two SU(2)$_L$ $\times$ U(1)$_Y$ multiplets $Q_L$ and $X_L$ which are in the \textbf{2}$_{1/6}$ and \textbf{2}$_{7/6}$ representations, respectively.  Since we are only working to leading order in $v^2/f^2$ we need only consider the mixing between the top multiplet and the composite multiplet with the same hypercharge, the $Q_L$.  The singlet $\Psi_1$ is typically written as $\tilde{U}_R$.  The mixing is
\begin{equation}
  \mathcal{L} \supset - m_4 \bar{Q}_L Q_L - m_1 \bar{\tilde{U}}_R \tilde{U}_R
  + y_L f (\bar{q}_L Q_L + \text{h.c.}) + y_R f (\bar{u}_R \tilde{U}_R + \text{h.c.}) .
\end{equation}
Let us rename the elementary fields with hats, $q_L \to \hat{q}_L$, $u_R \to \hat{u}_R$, $Q_L \to \hat{Q}_L$, and $\tilde{U}_R \to \hat{\tilde{U}}_R$.  The left handed quarks are then found with the rotation
\begin{equation}\label{eq:rot-L-fermion}
\begin{aligned}
  \hat{q}_L &= c_L^t q_L - s_L^t Q_L, \\
  \hat{Q}_L &= s_L^t q_L + c_L^t Q_L,
\end{aligned}
\quad\quad\quad\quad
s_L^t \equiv \frac{y_L}{\sqrt{(m_4/f)^2 + y_L^2}}, \quad
c_L^t \equiv \frac{m_4/f}{\sqrt{(m_4/f)^2 + y_L^2}}, 
\end{equation}
and similarly for the right handed quarks
\begin{equation}\label{eq:rot-R-fermion}
\begin{aligned}
  \hat{u}_R         &= c_R^t u_R - s_R^t \tilde{U}_R, \\
  \hat{\tilde{U}}_R &= s_R^t u_R + c_R^t \tilde{U}_R,
\end{aligned}
\quad\quad\quad\quad
s_R^t \equiv \frac{y_R}{\sqrt{(m_1/f)^2 + y_R^2}}, \quad
c_R^t \equiv \frac{m_1/f}{\sqrt{(m_1/f)^2 + y_R^2}}.
\end{equation}
In our simplified parameter space these are used as
\begin{equation}
  \sin\phi_{L,R} \equiv \frac{y_{L,R}}{\sqrt{(m_\Psi/f)^2 + y_{L,R}^2}} .
\end{equation}

\subsection{Vector interactions}\label{sec:interactions}

Having derived the rotations and specified the Lagrangian in Eq.~\eqref{eq:ele-comp-vectors}, we can now perform the rotations and derive the interactions in the physical basis (before electroweak symmetry breaking).

The interactions that emerge can be divided into two pieces: the SU(2)$_L$ triplet Lagrangian and the singlet Lagrangian ({\it i.e.} the SU(2)$_R$ triplet).  The triplet Lagrangian is \cite{Pappadopulo:2014qza}
\begin{equation}\label{eq:lag-left}
\begin{aligned}
\mathcal{L}_{\rm triplet} = &
- \frac{1}{4} D_{[\mu} \rho_{\nu]}^a D^{[\mu} \rho^{\nu]\;a}
+ \frac{m_\rho^2}{2} \; \rho^a_\mu \; \rho^{\mu a} \\
& + i g_\rho \bar{c}_H \rho^a_\mu H^\dagger \tau^a \overset{\leftrightarrow}{D}^\mu H
+ g_\rho^2 c_{\rho\rho HH} \rho_\mu^a \rho^{\mu a} H^\dagger H \\
& + \frac{g^2}{g_\rho} c_3 \rho_\mu^a J_3^{\mu a} 
+ \frac{g^2}{g_\rho} c_q \rho_\mu^a J_q^{\mu a} 
+ \frac{g^2}{g_\rho} c_\ell \rho_\mu^a J_\ell^{\mu a} \\
& - \frac{g}{2} c_{\rho\rho W} \epsilon_{abc} W^{\mu\nu a} \rho_\mu^b \rho_\nu^c
+ \frac{g_\rho}{2} c_{\rho\rho\rho} \epsilon_{abc} \rho_\mu^a \rho_\nu^b D^{[\mu} \rho^{\nu]\; c}
- \frac{g_\rho^2}{4} c_{\rho\rho\rho\rho} \epsilon_{abe} \epsilon_{cde} \rho_\mu^a \rho_\nu^b \rho^{\mu c} \rho^{\nu d}.
\end{aligned}
\end{equation}
The coefficients in our two-site model are
\begin{equation}
\begin{aligned}
\bar{c}_H            &= c_H + \order{\nicefrac{g^2}{g_\rho^2}} =\frac{1}{2} + \order{\nicefrac{g^2}{g_\rho^2}}, &
\quad\quad\quad\quad\quad
c_{\rho\rho HH}      &= \order{\nicefrac{g^2}{g_\rho^2}}, \\
c_3                  &= -(1-s_{L,t}^2 \nicefrac{g_\rho^2}{g^2}) + \order{\nicefrac{g^2}{g_\rho^2}}, &
c_{\rho\rho W}       &= 1, \\
c_q                  &= -(1-s_{L,q}^2 \nicefrac{g_\rho^2}{g^2}) + \order{\nicefrac{g^2}{g_\rho^2}}, &
c_{\rho\rho\rho}     &= -1 + \order{\nicefrac{g^2}{g_\rho^2}}, \\
c_\ell               &= -1 + \order{\nicefrac{g^2}{g_\rho^2}}, &
c_{\rho\rho\rho\rho} &= 1 +  \order{\nicefrac{g^2}{g_\rho^2}}.
\end{aligned}
\end{equation}
The covariant derivative on $\rho$ is defined as
\begin{equation}
D_{[\mu} \rho_{\nu]}^a \equiv D_\mu \rho_\nu^a - D_\nu \rho_\mu^a ,
\quad\quad\quad
D_\mu \rho_\nu^a \equiv \partial_\mu \rho_\nu^a + g \epsilon^{abc} W_\mu^b \rho_\nu^c ,
\end{equation}
and the operator $i g_\rho \bar c_H \rho_\mu^a H^\dagger \tau^a \overset{\leftrightarrow}{D}^\mu H$ is
\begin{equation}
i g_\rho \bar c_H \rho_\mu^a H^\dagger \tau^a \overset{\leftrightarrow}{D}^\mu H
= i g_\rho \bar c_H \rho_\mu^a (H^\dagger \tau^a D^\mu H - (D^\mu H)^\dagger \tau^a H).
\end{equation}
The fermion currents are defined as
\begin{subequations}
\begin{align}
J_3^{\mu a}    &= \sum_f \bar{f}_L \gamma^\mu \tau^a f_L ,   && f = \{b, t\}, \\
J_q^{\mu a}    &= \sum_f \bar{f}_L \gamma^\mu \tau^a f_L ,   && f = \{u, d, c, s\}, \\
J_\ell^{\mu a} &= \sum_f \bar{f}_L \gamma^\mu \tau^a f_L ,   && f = \{e, \mu, \tau, \nu_e, \nu_\mu, \nu_\tau \}.
\end{align}
\end{subequations}
The relevant singlet Lagrangian, which extends the triplet case of~\cite{Pappadopulo:2014qza}, is
\begin{equation}\label{eq:lag-b}
\begin{aligned}
\mathcal{L}_{\rm singlet} = &
- \frac{1}{4} \rho_{B\mu\nu} \rho_B^{\mu\nu} 
+ \frac{m_{\rho_B}^2}{2} \; \rho_{B\mu} \; \rho_B^\mu
+ i \frac{g_\rho}{2} \bar{c}_H^B \rho_{B\mu} H^\dagger \overset{\leftrightarrow}{D}^\mu H \\
&- \frac{1}{2} D_{[\mu} \rho^+_{C\nu]} D^{[\mu} \rho_C^{-\nu]}
+ m_{\rho_C}^2 \; \rho_{C\mu}^+ \; \rho_C^{-\mu} 
+ i g_\rho \bar{c}_H^C (H^\dagger \rho_C^{+\mu} \overset{\leftrightarrow}{D}_\mu H^c + H^{c\dagger} \rho_C^{-\mu} \overset{\leftrightarrow}{D}_\mu H) \\
& + \frac{g'^2}{g_\rho} c_3^{B,q_L} \rho_\mu^a J_{3,q_L}^\mu
+ \frac{g'^2}{g_\rho} c_3^{B,u_R} \rho_\mu^a J_{3,u_R}^\mu
+ \frac{g'^2}{g_\rho} c_3^{B,d_R} \rho_\mu^a J_{3,d_R}^\mu \\
& + \frac{g'^2}{g_\rho} c_q^{B,q_L} \rho_\mu^a J_{q,q_L}^\mu
+ \frac{g'^2}{g_\rho} c_q^{B,u_R} \rho_\mu^a J_{q,u_R}^\mu
+ \frac{g'^2}{g_\rho} c_q^{B,d_R} \rho_\mu^a J_{q,d_R}^\mu \\
& + \frac{g'^2}{g_\rho} c_\ell^{B,\ell_L} \rho_\mu^a J_{\ell,\ell_L}^\mu
+ \frac{g'^2}{g_\rho} c_\ell^{B,e_R} \rho_\mu^a J_{\ell,e_R}^\mu ,
\end{aligned}
\end{equation}
where we have omitted the interactions with more than one vector (which are similar to Eq.~\eqref{eq:lag-left}).  The coefficients are
\begin{equation}
\begin{aligned}
\bar{c}_H^B \quad    &= c_H + \order{\nicefrac{g'^2}{g_\rho^2}} =\frac{1}{2} + \order{\nicefrac{g'^2}{g_\rho^2}}, &
\quad\quad\quad\quad\quad
\bar{c}_H^C \quad    &= c_H + \order{\nicefrac{g'^2}{g_\rho^2}}  =\frac{1}{2} + \order{\nicefrac{g'^2}{g_\rho^2}}, \\
c_3^{B,q_L}          &= -\nicefrac{1}{6}(1-s_{L,t}^2 \nicefrac{g_\rho^2}{g'^2}) + \order{\nicefrac{g'^2}{g_\rho^2}}, &
c_q^{B,q_L}          &= -\nicefrac{1}{6}(1-s_{L,q}^2 \nicefrac{g_\rho^2}{g'^2}) + \order{\nicefrac{g'^2}{g_\rho^2}}, \\
c_3^{B,u_R}          &= -\nicefrac{2}{3} + \order{\nicefrac{g'^2}{g_\rho^2}}, &
c_q^{B,u_R}          &= -\nicefrac{2}{3} + \order{\nicefrac{g'^2}{g_\rho^2}}, \\
c_3^{B,d_R}          &= \nicefrac{1}{3} + \order{\nicefrac{g'^2}{g_\rho^2}}, &
c_q^{B,d_R}          &= \nicefrac{1}{3} + \order{\nicefrac{g'^2}{g_\rho^2}}, \\
c_3^{B,\ell_L}       &= \nicefrac{1}{2} + \order{\nicefrac{g'^2}{g_\rho^2}}, &
c_q^{B,e_R}          &= 1 + \order{\nicefrac{g'^2}{g_\rho^2}},
\end{aligned}
\end{equation}
and the fermion currents are
\begin{subequations}
\begin{align}
J^\mu_{3,f}    &= \sum_f \bar{f} \gamma^\mu f ,   && f = \{b, t\}, \\
J^\mu_{q,f}    &= \sum_f \bar{f} \gamma^\mu f ,   && f = \{u, d, c, s\}, \\
J^\mu_{\ell,f} &= \sum_f \bar{f} \gamma^\mu f ,   && f = \{e, \mu, \tau, \nu_e, \nu_\mu, \nu_\tau \}.
\end{align}
\end{subequations}

\pagestyle{plain}
\bibliographystyle{jhep}
\small\bibliography{biblio}

\providecommand{\href}[2]{#2}\begingroup\raggedright\begin{thebibliography}{10}

\bibitem{Aad:2015owa}
{\bf ATLAS} Collaboration, G.~Aad et~al., {\it {Search for high-mass diboson
  resonances with boson-tagged jets in proton-proton collisions at $\sqrt{s}$ =
  8 TeV with the ATLAS detector}},  \href{http://arxiv.org/abs/1506.00962}{{\tt
  arXiv:1506.00962}}.

\bibitem{Khachatryan:2014hpa}
{\bf CMS} Collaboration, V.~Khachatryan et~al., {\it {Search for massive
  resonances in dijet systems containing jets tagged as W or Z boson decays in
  pp collisions at $ \sqrt{s} $ = 8 TeV}},  {\em JHEP} {\bf 08} (2014) 173,
  [\href{http://arxiv.org/abs/1405.1994}{{\tt arXiv:1405.1994}}].

\bibitem{Aad:2014cka}
{\bf ATLAS} Collaboration, G.~Aad et~al., {\it {Search for high-mass dilepton
  resonances in pp collisions at $\sqrt{s}=8$  TeV with the ATLAS
  detector}},  {\em Phys.Rev.} {\bf D90} (2014), no.~5 052005,
  [\href{http://arxiv.org/abs/1405.4123}{{\tt arXiv:1405.4123}}].

\bibitem{Khachatryan:2014fba}
{\bf CMS} Collaboration, V.~Khachatryan et~al., {\it {Search for physics beyond
  the standard model in dilepton mass spectra in proton-proton collisions at $
  \sqrt{s}=8 $ TeV}},  {\em JHEP} {\bf 1504} (2015) 025,
  [\href{http://arxiv.org/abs/1412.6302}{{\tt arXiv:1412.6302}}].

\bibitem{Fukano:2015hga}
H.~S. Fukano, M.~Kurachi, S.~Matsuzaki, K.~Terashi, and K.~Yamawaki, {\it {2
  TeV Walking Technirho at LHC?}},  \href{http://arxiv.org/abs/1506.03751}{{\tt
  arXiv:1506.03751}}.

\bibitem{Franzosi:2015zra}
D.~B. Franzosi, M.~T. Frandsen, and F.~Sannino, {\it {Diboson Signals via Fermi
  Scale Spin-One States}},  \href{http://arxiv.org/abs/1506.04392}{{\tt
  arXiv:1506.04392}}.

\bibitem{Dobrescu:2015qna}
B.~A. Dobrescu and Z.~Liu, {\it {A W' boson near 2 TeV: predictions for Run 2
  of the LHC}},  \href{http://arxiv.org/abs/1506.06736}{{\tt
  arXiv:1506.06736}}.

\bibitem{Gao:2015irw}
Y.~Gao, T.~Ghosh, K.~Sinha, and J.-H. Yu, {\it {G221 Interpretations of the
  Diboson and Wh Excesses}},  \href{http://arxiv.org/abs/1506.07511}{{\tt
  arXiv:1506.07511}}.

\bibitem{Brehmer:2015cia}
J.~Brehmer, J.~Hewett, J.~Kopp, T.~Rizzo, and J.~Tattersall, {\it {Symmetry
  Restored in Dibosons at the LHC?}},
  \href{http://arxiv.org/abs/1507.00013}{{\tt arXiv:1507.00013}}.

\bibitem{Pappadopulo:2014qza}
D.~Pappadopulo, A.~Thamm, R.~Torre, and A.~Wulzer, {\it {Heavy Vector Triplets:
  Bridging Theory and Data}},  {\em JHEP} {\bf 1409} (2014) 060,
  [\href{http://arxiv.org/abs/1402.4431}{{\tt arXiv:1402.4431}}].

\bibitem{Aguilar-Saavedra:2015rna}
J.~Aguilar-Saavedra, {\it {Triboson interpretations of the ATLAS diboson
  excess}},  \href{http://arxiv.org/abs/1506.06739}{{\tt arXiv:1506.06739}}.

\bibitem{Thamm:2015csa}
A.~Thamm, R.~Torre, and A.~Wulzer, {\it {A composite Heavy Vector Triplet in
  the ATLAS di-boson excess}},  \href{http://arxiv.org/abs/1506.08688}{{\tt
  arXiv:1506.08688}}.

\bibitem{Sanz:2015zha}
V.~Sanz, {\it {On the compatibility of the diboson excess with a gg-initiated
  composite sector}},  \href{http://arxiv.org/abs/1507.03553}{{\tt
  arXiv:1507.03553}}.

\bibitem{Bian:2015ota}
L.~Bian, D.~Liu, and J.~Shu, {\it {Low Scale Composite Higgs Model and 1.8
  $\sim$ 2 TeV Diboson Excess}},  \href{http://arxiv.org/abs/1507.06018}{{\tt
  arXiv:1507.06018}}.

\bibitem{Carmona:2015xaa}
A.~Carmona, A.~Delgado, M.~Quiros, and J.~Santiago, {\it {Diboson resonant
  production in non-custodial composite Higgs models}},
  \href{http://arxiv.org/abs/1507.01914}{{\tt arXiv:1507.01914}}.

\bibitem{Hisano:2015gna}
J.~Hisano, N.~Nagata, and Y.~Omura, {\it {Interpretations of the ATLAS Diboson
  Resonances}},  \href{http://arxiv.org/abs/1506.03931}{{\tt
  arXiv:1506.03931}}.

\bibitem{Cheung:2015nha}
K.~Cheung, W.-Y. Keung, P.-Y. Tseng, and T.-C. Yuan, {\it {Interpretations of
  the ATLAS Diboson Anomaly}},  \href{http://arxiv.org/abs/1506.06064}{{\tt
  arXiv:1506.06064}}.

\bibitem{Xue:2015wha}
S.-S. Xue, {\it {Vector-like $W^\pm$-boson coupling at TeV and fermion-mass
  hierarchy (two boson-tagged jets vs four quark jets)}},
  \href{http://arxiv.org/abs/1506.05994}{{\tt arXiv:1506.05994}}.

\bibitem{Chao:2015eea}
W.~Chao, {\it {ATLAS Diboson Excesses from the Stealth Doublet Model}},
  \href{http://arxiv.org/abs/1507.05310}{{\tt arXiv:1507.05310}}.

\bibitem{Omura:2015nwa}
Y.~Omura, K.~Tobe, and K.~Tsumura, {\it {Survey of Higgs interpretations of the
  diboson excesses}},  \href{http://arxiv.org/abs/1507.05028}{{\tt
  arXiv:1507.05028}}.

\bibitem{Chen:2015xql}
C.-H. Chen and T.~Nomura, {\it {2 TeV Higgs boson and ATLAS diboson excess}},
  \href{http://arxiv.org/abs/1507.04431}{{\tt arXiv:1507.04431}}.

\bibitem{Chiang:2015lqa}
C.-W. Chiang, H.~Fukuda, K.~Harigaya, M.~Ibe, and T.~T. Yanagida, {\it {Diboson
  Resonance as a Portal to Hidden Strong Dynamics}},
  \href{http://arxiv.org/abs/1507.02483}{{\tt arXiv:1507.02483}}.

\bibitem{Cacciapaglia:2015nga}
G.~Cacciapaglia, A.~Deandrea, and M.~Hashimoto, {\it {A scalar hint from the
  diboson excess?}},  \href{http://arxiv.org/abs/1507.03098}{{\tt
  arXiv:1507.03098}}.

\bibitem{Alves:2015mua}
A.~Alves, A.~Berlin, S.~Profumo, and F.~S. Queiroz, {\it {Dirac-Fermionic Dark
  Matter in $U(1)\_X$ Models}},  \href{http://arxiv.org/abs/1506.06767}{{\tt
  arXiv:1506.06767}}.

\bibitem{Cao:2015lia}
Q.-H. Cao, B.~Yan, and D.-M. Zhang, {\it {Simple Non-Abelian Extensions and
  Diboson Excesses at the LHC}},  \href{http://arxiv.org/abs/1507.00268}{{\tt
  arXiv:1507.00268}}.

\bibitem{Abe:2015jra}
T.~Abe, R.~Nagai, S.~Okawa, and M.~Tanabashi, {\it {Unitarity sum rules, three
  site moose model, and the ATLAS 2 TeV diboson anomalies}},
  \href{http://arxiv.org/abs/1507.01185}{{\tt arXiv:1507.01185}}.

\bibitem{Anchordoqui:2015uea}
L.~A. Anchordoqui, I.~Antoniadis, H.~Goldberg, X.~Huang, D.~Lust, and T.~R.
  Taylor, {\it {Stringy origin of diboson and dijet excesses at the LHC}},
  \href{http://arxiv.org/abs/1507.05299}{{\tt arXiv:1507.05299}}.

\bibitem{Englert:2015oga}
C.~Englert, P.~Harris, M.~Spannowsky, and M.~Takeuchi, {\it
  {Unitarity-controlled resonances after the Higgs boson discovery}},  {\em
  Phys. Rev.} {\bf D92} (2015), no.~1 013003,
  [\href{http://arxiv.org/abs/1503.07459}{{\tt arXiv:1503.07459}}].

\bibitem{Contino:2010rs}
R.~Contino, {\it {The Higgs as a Composite Nambu-Goldstone Boson}},
  \href{http://arxiv.org/abs/1005.4269}{{\tt arXiv:1005.4269}}.

\bibitem{Panico:2015jxa}
G.~Panico and A.~Wulzer, {\it {The Composite Nambu-Goldstone Higgs}},
  \href{http://arxiv.org/abs/1506.01961}{{\tt arXiv:1506.01961}}.

\bibitem{ATLAS:2015bea}
{\bf ATLAS} Collaboration, T.~A. collaboration, {\it {Measurements of the Higgs
  boson production and decay rates and coupling strengths using pp collision
  data at √s = 7 and 8 TeV in the ATLAS experiment}}, .

\bibitem{Khachatryan:2014jba}
{\bf CMS} Collaboration, V.~Khachatryan et~al., {\it {Precise determination of
  the mass of the Higgs boson and tests of compatibility of its couplings with
  the standard model predictions using proton collisions at 7 and 8 $\,\text
  {TeV}$}},  {\em Eur. Phys. J.} {\bf C75} (2015), no.~5 212,
  [\href{http://arxiv.org/abs/1412.8662}{{\tt arXiv:1412.8662}}].

\bibitem{Kaplan:1991dc}
D.~B. Kaplan, {\it {Flavor at SSC energies: A New mechanism for
  dynamicallygenerated fermion masses}},  {\em Nucl. Phys.} {\bf B365} (1991)
  259--278.

\bibitem{Contino:2006qr}
R.~Contino, L.~Da~Rold, and A.~Pomarol, {\it {Light custodians in natural
  composite Higgs models}},  {\em Phys.Rev.} {\bf D75} (2007) 055014,
  [\href{http://arxiv.org/abs/hep-ph/0612048}{{\tt hep-ph/0612048}}].

\bibitem{Barducci:2012kk}
D.~Barducci, A.~Belyaev, S.~De~Curtis, S.~Moretti, and G.~M. Pruna, {\it
  {Exploring Drell-Yan signals from the 4D Composite Higgs Model at the LHC}},
  {\em JHEP} {\bf 1304} (2013) 152, [\href{http://arxiv.org/abs/1210.2927}{{\tt
  arXiv:1210.2927}}].

\bibitem{Bellazzini:2012tv}
B.~Bellazzini, C.~Csaki, J.~Hubisz, J.~Serra, and J.~Terning, {\it {Composite
  Higgs Sketch}},  {\em JHEP} {\bf 11} (2012) 003,
  [\href{http://arxiv.org/abs/1205.4032}{{\tt arXiv:1205.4032}}].

\bibitem{Vignaroli:2014bpa}
N.~Vignaroli, {\it {New W′ signals at the LHC}},  {\em Phys. Rev.} {\bf D89}
  (2014), no.~9 095027, [\href{http://arxiv.org/abs/1404.5558}{{\tt
  arXiv:1404.5558}}].

\bibitem{Greco:2014aza}
D.~Greco and D.~Liu, {\it {Hunting composite vector resonances at the LHC:
  naturalness facing data}},  {\em JHEP} {\bf 1412} (2014) 126,
  [\href{http://arxiv.org/abs/1410.2883}{{\tt arXiv:1410.2883}}].

\bibitem{Agashe:2007ki}
K.~Agashe, H.~Davoudiasl, S.~Gopalakrishna, T.~Han, G.-Y. Huang, G.~Perez,
  Z.-G. Si, and A.~Soni, {\it {LHC Signals for Warped Electroweak Neutral Gauge
  Bosons}},  {\em Phys. Rev.} {\bf D76} (2007) 115015,
  [\href{http://arxiv.org/abs/0709.0007}{{\tt arXiv:0709.0007}}].

\bibitem{Agashe:2008jb}
K.~Agashe, S.~Gopalakrishna, T.~Han, G.-Y. Huang, and A.~Soni, {\it {LHC
  Signals for Warped Electroweak Charged Gauge Bosons}},  {\em Phys. Rev.} {\bf
  D80} (2009) 075007, [\href{http://arxiv.org/abs/0810.1497}{{\tt
  arXiv:0810.1497}}].

\bibitem{Geller:2014kta}
M.~Geller and O.~Telem, {\it {Holographic Twin Higgs Model}},  {\em
  Phys.Rev.Lett.} {\bf 114} (2015), no.~19 191801,
  [\href{http://arxiv.org/abs/1411.2974}{{\tt arXiv:1411.2974}}].

\bibitem{Barbieri:2015lqa}
R.~Barbieri, D.~Greco, R.~Rattazzi, and A.~Wulzer, {\it {The Composite Twin
  Higgs scenario}},  \href{http://arxiv.org/abs/1501.07803}{{\tt
  arXiv:1501.07803}}.

\bibitem{Low:2015nqa}
M.~Low, A.~Tesi, and L.-T. Wang, {\it {Twin Higgs mechanism and a composite
  Higgs boson}},  {\em Phys.Rev.} {\bf D91} (2015) 095012,
  [\href{http://arxiv.org/abs/1501.07890}{{\tt arXiv:1501.07890}}].

\bibitem{Agashe:2004rs}
K.~Agashe, R.~Contino, and A.~Pomarol, {\it {The Minimal composite Higgs
  model}},  {\em Nucl.Phys.} {\bf B719} (2005) 165--187,
  [\href{http://arxiv.org/abs/hep-ph/0412089}{{\tt hep-ph/0412089}}].

\bibitem{Panico:2011pw}
G.~Panico and A.~Wulzer, {\it {The Discrete Composite Higgs Model}},  {\em
  JHEP} {\bf 1109} (2011) 135, [\href{http://arxiv.org/abs/1106.2719}{{\tt
  arXiv:1106.2719}}].

\bibitem{DeCurtis:2011yx}
S.~De~Curtis, M.~Redi, and A.~Tesi, {\it {The 4D Composite Higgs}},  {\em JHEP}
  {\bf 1204} (2012) 042, [\href{http://arxiv.org/abs/1110.1613}{{\tt
  arXiv:1110.1613}}].

\bibitem{Coleman:1969sm}
S.~R. Coleman, J.~Wess, and B.~Zumino, {\it {Structure of phenomenological
  Lagrangians. 1.}},  {\em Phys.Rev.} {\bf 177} (1969) 2239--2247.

\bibitem{Callan:1969sn}
J.~Callan, Curtis~G., S.~R. Coleman, J.~Wess, and B.~Zumino, {\it {Structure of
  phenomenological Lagrangians. 2.}},  {\em Phys.Rev.} {\bf 177} (1969)
  2247--2250.

\bibitem{Contino:2011np}
R.~Contino, D.~Marzocca, D.~Pappadopulo, and R.~Rattazzi, {\it {On the effect
  of resonances in composite Higgs phenomenology}},  {\em JHEP} {\bf 1110}
  (2011) 081, [\href{http://arxiv.org/abs/1109.1570}{{\tt arXiv:1109.1570}}].

\bibitem{DeSimone:2012fs}
A.~De~Simone, O.~Matsedonskyi, R.~Rattazzi, and A.~Wulzer, {\it {A First Top
  Partner Hunter's Guide}},  {\em JHEP} {\bf 1304} (2013) 004,
  [\href{http://arxiv.org/abs/1211.5663}{{\tt arXiv:1211.5663}}].

\bibitem{Giudice:2007fh}
G.~F. Giudice, C.~Grojean, A.~Pomarol, and R.~Rattazzi, {\it {The
  Strongly-Interacting Light Higgs}},  {\em JHEP} {\bf 06} (2007) 045,
  [\href{http://arxiv.org/abs/hep-ph/0703164}{{\tt hep-ph/0703164}}].

\bibitem{Falkowski:2011ua}
A.~Falkowski, C.~Grojean, A.~Kaminska, S.~Pokorski, and A.~Weiler, {\it {If no
  Higgs then what?}},  {\em JHEP} {\bf 11} (2011) 028,
  [\href{http://arxiv.org/abs/1108.1183}{{\tt arXiv:1108.1183}}].

\bibitem{Barbieri:2012tu}
R.~Barbieri, D.~Buttazzo, F.~Sala, D.~M. Straub, and A.~Tesi, {\it {A 125 GeV
  composite Higgs boson versus flavour and electroweak precision tests}},  {\em
  JHEP} {\bf 1305} (2013) 069, [\href{http://arxiv.org/abs/1211.5085}{{\tt
  arXiv:1211.5085}}].

\bibitem{Matsedonskyi:2014iha}
O.~Matsedonskyi, {\it {On Flavour and Naturalness of Composite Higgs Models}},
  {\em JHEP} {\bf 1502} (2015) 154, [\href{http://arxiv.org/abs/1411.4638}{{\tt
  arXiv:1411.4638}}].

\bibitem{Allanach:2015hba}
B.~C. Allanach, B.~Gripaios, and D.~Sutherland, {\it {Anatomy of the ATLAS
  diboson anomaly}},  \href{http://arxiv.org/abs/1507.01638}{{\tt
  arXiv:1507.01638}}.

\bibitem{Khachatryan:2014gha}
{\bf CMS} Collaboration, V.~Khachatryan et~al., {\it {Search for massive
  resonances decaying into pairs of boosted bosons in semi-leptonic final
  states at $\sqrt{s}=$ 8 TeV}},  {\em JHEP} {\bf 08} (2014) 174,
  [\href{http://arxiv.org/abs/1405.3447}{{\tt arXiv:1405.3447}}].

\bibitem{Aad:2014xka}
{\bf ATLAS} Collaboration, G.~Aad et~al., {\it {Search for resonant diboson
  production in the $\mathrm{\ell \ell }q\bar{q}$ final state in $pp$
  collisions at $\sqrt{s} = 8$ TeV with the ATLAS detector}},  {\em Eur. Phys.
  J.} {\bf C75} (2015), no.~2 69, [\href{http://arxiv.org/abs/1409.6190}{{\tt
  arXiv:1409.6190}}].

\bibitem{Aad:2015ufa}
{\bf ATLAS} Collaboration, G.~Aad et~al., {\it {Search for production of
  $WW/WZ$ resonances decaying to a lepton, neutrino and jets in $pp$ collisions
  at $\sqrt{s}=8$ TeV with the ATLAS detector}},  {\em Eur.Phys.J.} {\bf C75}
  (2015), no.~5 209, [\href{http://arxiv.org/abs/1503.04677}{{\tt
  arXiv:1503.04677}}].

\bibitem{Aad:2014pha}
{\bf ATLAS} Collaboration, G.~Aad et~al., {\it {Search for $WZ$ resonances in
  the fully leptonic channel using $pp$ collisions at $\sqrt{s}$ = 8 TeV with
  the ATLAS detector}},  {\em Phys.Lett.} {\bf B737} (2014) 223--243,
  [\href{http://arxiv.org/abs/1406.4456}{{\tt arXiv:1406.4456}}].

\bibitem{Khachatryan:2014xja}
{\bf CMS} Collaboration, V.~Khachatryan et~al., {\it {Search for new resonances
  decaying via WZ to leptons in proton-proton collisions at $\sqrt s =$ 8
  TeV}},  {\em Phys. Lett.} {\bf B740} (2015) 83--104,
  [\href{http://arxiv.org/abs/1407.3476}{{\tt arXiv:1407.3476}}].

\bibitem{Khachatryan:2015ywa}
{\bf CMS} Collaboration, V.~Khachatryan et~al., {\it {Search for narrow
  high-mass resonances in proton–proton collisions at $\sqrt{s}$ = 8 TeV
  decaying to a Z and a Higgs boson}},  {\em Phys. Lett.} {\bf B748} (2015)
  255--277, [\href{http://arxiv.org/abs/1502.04994}{{\tt arXiv:1502.04994}}].

\bibitem{Khachatryan:2015bma}
{\bf CMS} Collaboration, V.~Khachatryan et~al., {\it {Search for A Massive
  Resonance Decaying into a Higgs Boson and a W or Z Boson in Hadronic Final
  States in Proton-Proton Collisions at $\sqrt{s}$ = 8 TeV}},
  \href{http://arxiv.org/abs/1506.01443}{{\tt arXiv:1506.01443}}.

\bibitem{CMS:2015gla}
{\bf CMS} Collaboration, C.~Collaboration, {\it {Search for massive WH
  resonances decaying to $\ell \nu {\rm b \bar{b}}$ final state in the boosted
  regime at $\sqrt{s}=8$\,TeV}}, .

\bibitem{Aad:2015yza}
{\bf ATLAS} Collaboration, G.~Aad et~al., {\it {Search for a new resonance
  decaying to a W or Z boson and a Higgs boson in the $\ell \ell / \ell \nu /
  \nu \nu + b \bar{b}$ final states with the ATLAS detector}},  {\em Eur. Phys.
  J.} {\bf C75} (2015), no.~6 263, [\href{http://arxiv.org/abs/1503.08089}{{\tt
  arXiv:1503.08089}}].

\bibitem{Alwall:2014hca}
J.~Alwall, R.~Frederix, S.~Frixione, V.~Hirschi, F.~Maltoni, O.~Mattelaer,
  H.~S. Shao, T.~Stelzer, P.~Torrielli, and M.~Zaro, {\it {The automated
  computation of tree-level and next-to-leading order differential cross
  sections, and their matching to parton shower simulations}},  {\em JHEP} {\bf
  07} (2014) 079, [\href{http://arxiv.org/abs/1405.0301}{{\tt
  arXiv:1405.0301}}].

\bibitem{ATLAS:2014wra}
{\bf ATLAS} Collaboration, G.~Aad et~al., {\it {Search for new particles in
  events with one lepton and missing transverse momentum in $pp$ collisions at
  $\sqrt{s}$ = 8 TeV with the ATLAS detector}},  {\em JHEP} {\bf 1409} (2014)
  037, [\href{http://arxiv.org/abs/1407.7494}{{\tt arXiv:1407.7494}}].

\bibitem{Khachatryan:2014tva}
{\bf CMS} Collaboration, V.~Khachatryan et~al., {\it {Search for physics beyond
  the standard model in final states with a lepton and missing transverse
  energy in proton-proton collisions at sqrt(s) = 8 TeV}},  {\em Phys.Rev.}
  {\bf D91} (2015), no.~9 092005, [\href{http://arxiv.org/abs/1408.2745}{{\tt
  arXiv:1408.2745}}].

\bibitem{Aad:2014xea}
{\bf ATLAS} Collaboration, G.~Aad et~al., {\it {Search for $W' \to t\bar{b}$ in
  the lepton plus jets final state in proton-proton collisions at a
  centre-of-mass energy of $\sqrt{s}$ = 8 TeV with the ATLAS detector}},  {\em
  Phys. Lett.} {\bf B743} (2015) 235--255,
  [\href{http://arxiv.org/abs/1410.4103}{{\tt arXiv:1410.4103}}].

\bibitem{Chatrchyan:2014koa}
{\bf CMS} Collaboration, S.~Chatrchyan et~al., {\it {Search for W' $\to $ tb
  decays in the lepton + jets final state in pp collisions at $\sqrt{s}$ = 8
  TeV}},  {\em JHEP} {\bf 05} (2014) 108,
  [\href{http://arxiv.org/abs/1402.2176}{{\tt arXiv:1402.2176}}].

\bibitem{Aad:2015fna}
{\bf ATLAS} Collaboration, G.~Aad et~al., {\it {A search for $t\bar{t}$
  resonances using lepton-plus-jets events in proton-proton collisions at
  $\sqrt{s} = 8$ TeV with the ATLAS detector}},
  \href{http://arxiv.org/abs/1505.07018}{{\tt arXiv:1505.07018}}.

\bibitem{Khachatryan:2015sma}
{\bf CMS} Collaboration, V.~Khachatryan et~al., {\it {Search for resonant
  $\mathrm{t\bar{t}}$ production in proton-proton collisions at $\sqrt{s}$ = 8
  TeV}},  \href{http://arxiv.org/abs/1506.03062}{{\tt arXiv:1506.03062}}.

\bibitem{Aad:2014aqa}
{\bf ATLAS} Collaboration, G.~Aad et~al., {\it {Search for new phenomena in the
  dijet mass distribution using $p-p$ collision data at $\sqrt{s}=8$ TeV with
  the ATLAS detector}},  {\em Phys.Rev.} {\bf D91} (2015), no.~5 052007,
  [\href{http://arxiv.org/abs/1407.1376}{{\tt arXiv:1407.1376}}].

\bibitem{Khachatryan:2015sja}
{\bf CMS} Collaboration, V.~Khachatryan et~al., {\it {Search for resonances and
  quantum black holes using dijet mass spectra in proton-proton collisions at
  $\sqrt{s} =$ 8 TeV}},  {\em Phys.Rev.} {\bf D91} (2015), no.~5 052009,
  [\href{http://arxiv.org/abs/1501.04198}{{\tt arXiv:1501.04198}}].

\bibitem{Thamm:2015zwa}
A.~Thamm, R.~Torre, and A.~Wulzer, {\it {Future tests of Higgs compositeness:
  direct vs indirect}},  \href{http://arxiv.org/abs/1502.01701}{{\tt
  arXiv:1502.01701}}.

\bibitem{Baak:2014ora}
{\bf Gfitter Group} Collaboration, M.~Baak, J.~Cúth, J.~Haller, A.~Hoecker,
  R.~Kogler, K.~Mönig, M.~Schott, and J.~Stelzer, {\it {The global electroweak
  fit at NNLO and prospects for the LHC and ILC}},  {\em Eur. Phys. J.} {\bf
  C74} (2014) 3046, [\href{http://arxiv.org/abs/1407.3792}{{\tt
  arXiv:1407.3792}}].

\bibitem{Barbieri:2007bh}
R.~Barbieri, B.~Bellazzini, V.~S. Rychkov, and A.~Varagnolo, {\it {The Higgs
  boson from an extended symmetry}},  {\em Phys. Rev.} {\bf D76} (2007) 115008,
  [\href{http://arxiv.org/abs/0706.0432}{{\tt arXiv:0706.0432}}].

\bibitem{Contino:2015mha}
R.~Contino and M.~Salvarezza, {\it {One-loop effects from spin-1 resonances in
  Composite Higgs models}},  {\em JHEP} {\bf 07} (2015) 065,
  [\href{http://arxiv.org/abs/1504.02750}{{\tt arXiv:1504.02750}}].

\bibitem{Agashe:2006at}
K.~Agashe, R.~Contino, L.~Da~Rold, and A.~Pomarol, {\it {A Custodial symmetry
  for Zb anti-b}},  {\em Phys. Lett.} {\bf B641} (2006) 62--66,
  [\href{http://arxiv.org/abs/hep-ph/0605341}{{\tt hep-ph/0605341}}].

\bibitem{Redi:2011zi}
M.~Redi and A.~Weiler, {\it {Flavor and CP Invariant Composite Higgs Models}},
  {\em JHEP} {\bf 1111} (2011) 108, [\href{http://arxiv.org/abs/1106.6357}{{\tt
  arXiv:1106.6357}}].

\bibitem{Khachatryan:2014cja}
{\bf CMS} Collaboration, V.~Khachatryan et~al., {\it {Search for quark contact
  interactions and extra spatial dimensions using dijet angular distributions
  in proton–proton collisions at $\sqrt s =$ 8 TeV}},  {\em Phys. Lett.} {\bf
  B746} (2015) 79--99, [\href{http://arxiv.org/abs/1411.2646}{{\tt
  arXiv:1411.2646}}].

\bibitem{Aad:2015eha}
{\bf ATLAS} Collaboration, G.~Aad et~al., {\it {Search for New Phenomena in
  Dijet Angular Distributions in Proton-Proton Collisions at $\sqrt{s} = 8$ TeV
  Measured with the ATLAS Detector}},  {\em Phys. Rev. Lett.} {\bf 114} (2015),
  no.~22 221802, [\href{http://arxiv.org/abs/1504.00357}{{\tt
  arXiv:1504.00357}}].

\bibitem{Domenech:2012ai}
O.~Domenech, A.~Pomarol, and J.~Serra, {\it {Probing the SM with Dijets at the
  LHC}},  {\em Phys. Rev.} {\bf D85} (2012) 074030,
  [\href{http://arxiv.org/abs/1201.6510}{{\tt arXiv:1201.6510}}].

\bibitem{Abe:2015uaa}
T.~Abe, T.~Kitahara, and M.~M. Nojiri, {\it {Prospects for Spin-1 Resonance
  Search at 13 TeV LHC and the ATLAS Diboson Excess}},
  \href{http://arxiv.org/abs/1507.01681}{{\tt arXiv:1507.01681}}.

\bibitem{SalamWeiler}
G.~Salam and A.~Weiler, {\it {Collider Reach ($\beta$)}},  2014.
\newblock
  \href{http://collider-reach.web.cern.ch/}{http://collider-reach.web.cern.ch/}.

\end{thebibliography}\endgroup
\end{document}